\documentclass[11pt]{article}
\usepackage{latexsym,graphics,graphicx,amssymb,amsmath, amstext, amsfonts, amsthm, pifont, color,subfig,epsfig,times}
\usepackage{hyperref}
\newtheorem{defn}{Definition}

\newtheorem{note}{Remark}

\newcommand{\bc}{\begin{center}}
\newcommand{\ec}{\end{center}}
\newcommand{\bfl}{\begin{flushleft}}
\newcommand{\efl}{\end{flushleft}}

\newcommand{\beqa}{\begin{eqnarray}}
\newcommand{\eeqa}{\end{eqnarray}}
\newcommand{\ach}{\text{ach}}
\newcommand{\org}{\text{original}}

\newcommand{\poly}{\text{poly}}

\newcommand{\beqan}{\begin{eqnarray*}}
\newcommand{\eeqan}{\end{eqnarray*}}
\newcommand{\beq}{\begin{equation}}
\newcommand{\eeq}{\end{equation}}
\newcommand{\beit}{\begin{itemize}}
\newcommand{\eeit}{\end{itemize}}

\newcommand{\lbr}{\left \{ }
\newcommand{\rbr}{\right \} }

\newcommand{\SNR}{{\tt SNR}}
\newcommand{\mc}{\mathcal}
\newcommand{\mb}{\mathbb}

\newcommand{\bit}{\begin{itemize}}
\newcommand{\eit}{\end{itemize}}
\newcommand{\ben}{\begin{enumerate}}
\newcommand{\een}{\end{enumerate}}

\newcommand{\bdefn}{\begin{defn}}
\newcommand{\edefn}{\end{defn}}
\newcommand{\bnote}{\begin{note}}
\newcommand{\enote}{\end{note}}

\newcommand{\blem}{\begin{lemma}}
\newcommand{\elem}{\end{lemma}}
\newcommand{\bthm}{\begin{theorem}}
\newcommand{\ethm}{\end{theorem}}
\newcommand{\bpf}{\begin{proof}}
\newcommand{\epf}{\end{proof}}
\newcommand{\bcor}{\begin{corollary}}
\newcommand{\ecor}{\end{corollary}}
\newcommand{\bprop}{\begin{proposition}}
\newcommand{\eprop}{\end{proposition}}

\newtheorem{proposition}{Proposition}
\newtheorem{lemma}{Lemma}
\newtheorem{theorem}{Theorem}

\newtheorem{corollary}{Corollary}

\newcommand{\rvin}{\rho_v^{{-}}}
\newcommand{\rvout}{\rho_v^{{+}}}
\newcommand{\In}{{\text{In}}}
\newcommand{\Out}{{\text{Out}}}

\newcommand{\bigO}{\mathcal{O}}

\newcommand{\g}{\text{g}}
\newcommand{\cut}{\text{cut}}

\newcommand{\lpr}{ \left( }
\newcommand{\rpr}{ \right) }

\makeatletter

\author{Sreeram Kannan\thanks{Supported in part by NSF grants CCF 1017430 and
CNS 0721652.}  \ and Pramod Viswanath\thanks{Supported in part by NSF grants CCF 1017430 and
CNS 0721652.} }
\date{Coordinated Science Laboratory and Dept. of ECE\\
University of Illinois, Urbana-Champaign, IL 618l01\\
Email: {\tt \{kannan1,  pramodv\}@illinois.edu}}

\title{Capacity of Multiple Unicast in Wireless Networks: A Polymatroidal Approach}

\begin{document}
\maketitle

\begin{abstract}

A classical result in undirected wireline networks is the near optimality of routing (flow) for multiple-unicast traffic (multiple sources communicating independent messages to multiple destinations): the min cut upper bound is within a logarithmic factor of the number of sources of the max flow. In this paper we ``extend" the wireline result to the wireless context.

Our main result is the approximate optimality of a simple layering principle: {\em local physical-layer schemes combined with global routing}. We use the {\em reciprocity} of the wireless channel critically in this result.  Our formal result is in the context of channel models for which ``good" local schemes, that achieve the cut-set bound, exist (such as  Gaussian MAC and broadcast channels, broadcast erasure networks,  fast fading Gaussian networks).

Layered architectures, common in the engineering-design of wireless networks, can have near-optimal performance if the {\em locality} over which physical-layer schemes should operate is carefully designed. Feedback is shown to play a critical role in enabling the separation between the physical and the network layers. The key technical idea is the modeling of a wireless network by an undirected ``polymatroidal'' network, for which we establish a max-flow min-cut approximation theorem.
\end{abstract}

\section{Introduction}\label{sec:intro}

Wireless networks are typically engineered using a layered approach: the physical layer deals with the channel noise, the medium access control layer deals with scheduling of users in the wireless context, the network layer handles routing of information and the transport layer deals with network congestion. While this design methodology has several engineering advantages that have led to the proliferation of wireless networks, a fundamental understanding of layering architectures is still lacking.

In this paper, we look explicitly for layered communication strategies that are near optimal for multiple unicast in general wireless networks. This would serve two (distinct) objectives:
\beit \item To obtain the approximate capacity of multiple unicast in wireless networks, and
\item To establish a layered communication architecture that can guide engineering design. \eeit

\subsection{Prior Work}

Fundamental understanding of layering architectures has recently received plenty of attention from the networking community \cite{Chiang} \cite{SrikantNUM}, and scenarios have been identified under which a joint optimization of the transport and network layers naturally decompose into separate optimization problems, thus yielding a justification for the layered architecture. While there have been attempts to include certain aspects of the wireless medium into this framework \cite{WirelessNUM}, the understanding is far from complete.  In this paper, we take a fundamental, information theoretic perspective,  on if and when, the physical, medium access and network layers can be separately designed.

\subsubsection{Capacity Results for Wireless Networks}
Substantial progress has been made in the recent past in understanding the key aspects of  the wireless medium (broadcast and superposition) from an information-theoretic view point. In particular, the capacity of MIMO broadcast channel has been resolved \cite{MIMOBC}, approximate capacity of the $2$-user interference channel has been established \cite{ETW} and the approximate capacity of $2$-user $X$-channels \cite{MMK} \cite{JafShamaiX}, $K$-user interference channels with diversity \cite{CadJafIA}, \cite{ErgodicIA} \cite{RealIA2,RealIA1} has been obtained. While these results establish information theoretic understanding of several important (``physical layer") channels, there is no conceptual guideline to fit the  solutions for reliable communication for the channel in the context of a bigger network it could be a part of. In a different direction, there has been significant progress in understanding network-level capacity issues in the context of simple traffic models, starting from the breakthrough work \cite{ADT}, where the approximate capacity of single unicast is characterized, and later generalized to several scenarios: the approximate capacity of unicast in discrete memoryless networks is characterized in \cite{NoisyNetCod},  a separation result between analog and digital components in relay networks is established in \cite{AnandKumar} and the approximate capacity of broadcasting in Gaussian networks is established in \cite{KannanRV10}.

While unicast traffic in general Gaussian networks and multiple unicast traffic in single-hop Gaussian networks are reasonably well understood, the capacity of multiple unicast traffic in Gaussian wireless networks remains an open problem in multi-terminal information theory. In recent times, several research groups have made progress on this problem \cite{Ave2Unicast,DiggInt,ChungJafarMultihop,AveSab,ReddyVeeravalli,GunYenGoldPoor} , but the general problem still remains unsolved. Specific directions, with promise of success, involve simplifying the problem
by considering specific traffic patterns such as $2$-unicast
\cite{DiggInt,Ave2Unicast, Wang2, WangTse, WangTse2, KamAnaTse, SuhTse,JafarChung2X2X2, VazVar1}; another
approach is to consider more specific network topologies, like for
example, $K$ sources communicating to $K$ sinks via $L$ fully-connected layers of $K$
relays each \cite{ChungJafarMultihop,NiesenNazerWhiting}. While these existing works attempt to compute the degrees-of-freedom (or approximate capacity) exactly for specific instances of the problem, we adopt a different viewpoint and focus our attention on obtaining general results for arbitrary networks (at the expense of obtaining potentially weaker approximation in specific instances).

In the context of multiple-unicast in large wireless networks, there has been significant progress in understanding {\em scaling laws} for geographical wireless networks; beginning with the seminal work in \cite{GuptaKumar} and culminating in the
hierarchical relaying scheme in \cite{OzgLevTse}  and a combination of the two \cite{Niesen1, Niesen2, Ozgur1} (with several critical works in between \cite{Xie1, Lev1, Aeron, Xue,Frans}). Despite its significant advantages, the performance guarantees are only in the context of certain specific wireless network models and, more importantly, the communication scheme is not a representation of  a simple layered architecture for communication.

\subsubsection{Information-theoretic Layering Architectures}
Separation theorems form a basic tool in information theory: in his celebrated paper \cite{Shannon}, Shannon showed that source coding (compression) and channel coding (communication) can be separated without loss of optimality. Following this, several separation (and non-separation) theorems have been proved in the multi-terminal context (see, for example, \cite{SepHassibi,SourceChannelSeparation}). The result most relevant to the current discussion is the separation between network coding and channel coding proved in the pioneering work  in \cite{KotMedEff}. There it is shown that, for a wireline network composed of {\em independent noisy channels}, a separation architecture composed of a physical layer that performs independent coding for each channel and a network layer which transports bits across the induced noiseless network, is optimal. This is a very interesting structural result that holds under arbitrary traffic models and is proved without the necessity to compute the capacity of the network. Thus the question of studying the capacity of wireline networks can be reduced to the question of studying the capacity of capacitated {\em graphs}.

For $k$-unicast in wireline graphs, a very interesting dichotomy is known in the theoretical computer science literature:  for undirected graphs, the classical work of Leighton and Rao \cite{LeightonRao} shows that routing achieves the min-cut to within a $\bigO(\log k)$ factor and furthermore, there is a standing conjecture \cite{LiLi} \cite{HarKleinLehman} \cite{KramerSavari} \cite{Langberg} which claims that routing is infact optimal; whereas for general directed graphs, it has been shown recently \cite{ChuzhoyK07} that there is no polynomial time algorithm that can approximate the value of min-cut to within a $k^{\epsilon}$ factor. Since max-flow can be computed in polynomial time, this result implies that there are networks for which flow cut gap is greater than $k^{\epsilon}$. Furthermore, it has been proved recently \cite{ChanGrant} that computing the network coding region for directed graphs is equivalent to determining the entropy vector region, which is believed to be a very hard problem. Thus the string of positive results in the context of undirected (or bidirected) graphs and negative results in the context of directed graphs serve as an indicator that it may be easier to understand bidirected wireless networks.

 A study of layering in directed wireless networks is intiated in  \cite{KotMedEff2}. The key idea is that of channel emulation, where a given channel is upper bounded by a wireline network with joint capacity constraints in such a way that the wireline network can emulate all possible behavior of the channel. This is a very strong condition, which ensures that the channel can be upper bounded by the wireline network {\em irrespective of the traffic pattern}. This program has been already accomplished for networks of 2-user MAC and broadcast channels, but appears to be very hard for general networks. In this paper, we instead work with only multiple-unicast traffic but go on to study layering in general bidirected networks.

\subsubsection{Polymatroidal Networks}
 Polymatroidal networks and approximation theorems for these networks form a crictical component in this paper. Polymatroidal networks are similar to edge capacitated networks, however, in addition to having edge capacity constraints, there are joint capacity constraints on the set of edges which meet at a vertex. Polymatroidal networks
  were introduced by Lawler and Martel \cite{LawlerMartel} and Hassin
  \cite{Hassin} in the single-commodity setting and are closely
  related to the submodular flow model of Edmonds and Giles
  \cite{EdmondsGiles}; the well-known maxflow-mincut theorem
  holds in this more general setting (see Chapter~$60$ in \cite{Schrijver} for a discussion). Multiple unicast in polymatroidal networks has not been studied till the recent work of \cite{CKRV}, where an approximate max-flow min-cut result is established for undirected polymatroidal networks.

Recently, there have been several applications of network flows, and their generalizations such as flows in linking systems \cite{Schrijver-Thesis}, to information flow in wireless networks
\cite{ADT,AmaFrag,YazdiSavari,GoemansIZ,RajaV11,KimMed}. The polymatroidal structure of the multiple access channel capacity region was observed and exploited by Tse and Hanly \cite{TH}. Directed polymatroidal networks were utlized in the work of Vasudevan and Korada \cite{Korada}, where a separation architecture for a network of deterministic broadcast and MAC channels converts the network into a polymatroidal network and existing results for broadcasting in polymatroidal networks \cite{polymatroidBC} are used to obtain capacity bounds.

\subsection{Summary of Results}
An important contribution of this paper is the following {\em meta-theorem}:
\begin{quote}
{\bf Meta Theorem:} {\em If there is a ``good'' physical-layer scheme (which is approximately cut-set achieving and reciprocal) for a certain channel, then, for multiple-unicast in a network composed of such channels, a layered architecture is approximately cut-set achieving (to within a logarithmic factor in the number of messages). }
\end{quote}
\begin{figure}[htb]
\begin{centering}
\scalebox{0.6}{\includegraphics{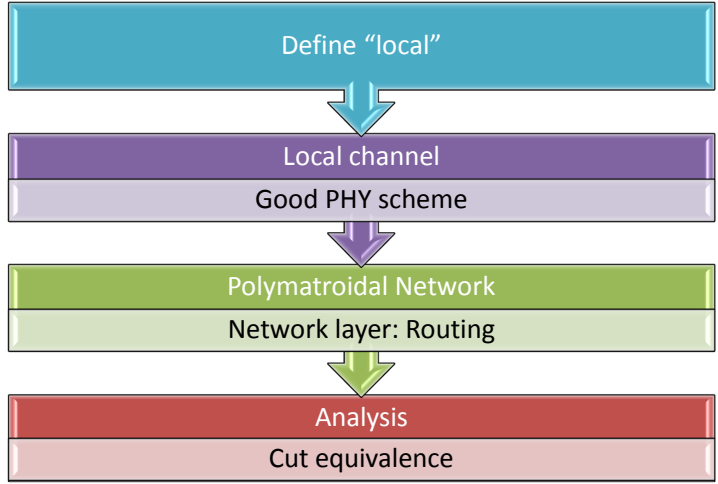}}
\caption{Layered Architecture} \label{fig:lay}
\end{centering}
\end{figure}

Fig.~\ref{fig:lay} depicts the layered architecture used in this paper. The following is a summary of the key ideas which are used in this paper to argue that a layered architecture is approximately capacity optimal:

\beit \item Model a wireless network as a bidirected network, by using the natural reciprocity of wireless networks.
\item Utilize a good local ``physical layer'' scheme for each channel and identify the combinatorial structure of the rate region (typically submodularity).
\item Show that local physical layer schemes convert a wireless network into a {\em bidirected polymatroidal network}. Thus the bidirected polymatroidal network can be viewed as a {\em graphical model} for wireless networks.
\item Prove a Leighton-Rao type approximation result for bidirected polymatroidal network, which shows that routing is near optimal for $k$-unicast traffic.
\item Argue that the layered architecture with local physical layer scheme + global routing achieves the cut-set approximately in the wireless network.
\item We provide a technique by which ``good'' results for a given channel can be {\em lifted up} to good results for a general network comprised of those channels.
\eeit

We justify the meta-thereom formally in the context of the following channel models:
\ben \item Networks composed of Gaussian broadcast and MAC channels
\item Networks composed of broadcast erasure channels with feedback
\item Fast fading wireless networks
\item Degrees-of-freedom approximation for fixed wireless networks
\item Linear deterministic networks composed of MAC and broadcast channels
\item Networks composed of MIMO MAC and broadcast channels with delayed CSI feedback
\item Fast fading linear deterministic networks
\een

For each of these networks, under a general $k$-unicast traffic model, the approximation factor on the rate is $\bigO(\log k)$ for the {\em entire rate region} in addition to the loss incurred due to the physical layer scheme, which is typically a power-scaling loss. Under more specific traffic models, such as the $X$-traffic model (where each of the $J$ sources have messages to send to each of the $K$ destinations) or a ``group-communication'' traffic model (where a subset $S$ of nodes have messages to send to each other), we prove a constant approximation factor for the {\em sum-rate}, again in addition to a power scaling loss (the constant being $1$, $2$ or $4$ depending on the specific channel and traffic model).

The material in this paper was presented in part in \cite{KannanRV11,KannanV11}.

\subsection{Organization}
The rest of this document is organized as follows:
\beit \item  An overview of the layering approach is provided in  Sec.~\ref{sec:layered}. After defining the ``locality'' over which local physical layer schemes must be implemented, a list of desirable properties of local solutions are provided.
\item In Sec.~\ref{sec:poly}, polymatroidal networks, which form the back-bone of the layering architecture, are defined and results for multiple-unicast in polymatroidal networks are presented.
\item In Sec.~\ref{sec:local_schemes}, ``good'' local physical layer solutions are described for various channel models. For some models, it is shown that existing schemes satisfy the desirable properties, whereas for other models, where existing schemes are insufficient, new ones are constructed.
\item In Sec.~\ref{sec:approx_capacity}, the local schemes are fitted into a global network context. Capacity theorems are proved for the various channel settings by connecting the wireless network problem formally to the polymatroidal network problem.
\eeit

\section{Layered Architecture \label{sec:layered}}
Engineering approaches to reliable network communication involve
``layering", a separation of the roles of physical (dealing with
channel uncertainty), medium access (dealing with sharing the wireless
medium) and networking (dealing with end-to-end the resulting
``wireline" network communication). On the other hand, fundamental
architectures are suggested by information theoretic study of large
wireless networks (a major research direction in the past decade, with
performance measured in a coarse scaling context). For instance,
multihop routing \cite{GuptaKumar} is a layered architecture, while hierarchical MIMO \cite{OzgLevTse}
(nearly scaling-law optimal in a geographically uniform context) is
not. The information theoretic understanding of layering architectures has recently started receiving attention (see \cite{KotMedEff,KotMedEff2}). Our approach is in similar lines as the approach in \cite{Korada}, where a layered architecture for a network of deterministic broadcast and MAC channels is used to obtain capacity bounds.

In understanding the systematic design of layered architectures, it helps to look at the global wireless network as a collection of ``local" wireless networks. The focus of this section is to introduce
this view point; we propose that the notion of locality comes from both geographic (spatial) and temporal contexts. We see that certain combinatorial properties of the (physical layer) solutions to the local networks are {\em desirable}; these will help prove fundamental guarantees on the performance of the layered architecture in the global context.

\subsection{Locality}
A wireless network is a collection of {\em local} channels, if there are no interactions between the channels. Formally, a wireless network composed is defined as follows: Consider a graph $G = (V,E)$. For each $v \in V$, $x_v,y_v$ denote the transmit and received symbol respectively. Let $C^{+}(v),C^{-}(v)$ denote the set of all channels in which $v$ can transmit and receive from respectively, i.e., $x_v$ can be written as $x_v = \{x_v^c\}_{c \in C^{+}(v)}$, and $y_v = \{y_v^c\}_{c \in C^{-}(v)}$. We will consider a wireless network with independent noise, where
\beqa \mathbb{P}(y_1,\ldots ,y_n | x_1,\ldots ,x_n) = \Pi_v \mathbb{P}( y_v | \{x_u\}_{u:(u,v) \in E}),\eeqa
and this description explicitly captures the relationship between the graph and the joint probability transition function.

Consider a set of channels $c \in \mc{C}$. A wireless network  is said to be composed of channels $c \in \mc{C}$ if the probabilistic description of the network is of the form,
\beqa \mathbb{P}(y_1,\ldots ,y_n | x_1,\ldots ,x_n) & = & \Pi_c \mathbb{P}( \{y_v^c\}_{v \in V^{-}(c)} | \{x_u^c\}_{u \in V^{+}(c)}) \\
& = &  \Pi_c \Pi_{v \in V^{-}(c)} \mathbb{P}( y^c_v | \{x_u^c\}_{u \in V^{+}(c): (u,v) \in E}),\eeqa where $V^{-}(c) = \{ v: c \in C^{-}(v) \}$ and $V^{+}(c) = \{ v: c \in C^{+}(v) \}$. Each channel $c$ is referred to as a component channel of the network.

A canonical scenario occurs when the wireless network is simply a collection of statistically independent noisy channels. Here each channel between a transmitter and a receiver is local. A more interesting example occurs in the case of a {\em frequency planned} wireless network, where each component of the wireless network operates in a specified frequency range. Here, the overall channel model can be decomposed as the product of channel models in each frequency range; the scale of locality corresponds to the scale of frequency reuse.

In general, such a geographic decomposition (via frequency planning) may not happen. Nevertheless, we can view the decomposition as occurring in {\em time} (indeed, this has been a popular method for analyzing general wireline / wireless networks \cite{NetCod,ADT}).  When we decompose across time, the local channel corresponds to the global one, as viewed over a specific single block of time. In this context, the layering architecture restricts the sophistication of physical layer (and medium access layer) strategies to be  restricted to operate on a single layer in time, and at the end of each epoch, the information is {\em decoded} and re-encoded (using the networking layer) for the next local channel. The layering architecture thus enforces decoding of all information at each ``hop" (in time); schemes, such as quantize-and-forward \cite{ADT}, which forward analog information do not fit the layered architecture model.

\subsection{Desirable Properties of Local Solutions}
A natural desirable property of any (physical layer) solution to a local channel is it be as optimal (from an information theoretic view point) as possible. In particular, we will be interested in how close the solution is to fundamental upper bounds given by the cutset bounds and certain natural combinatorial properties of the solution. For a network described by a probability transition matrix, the cut-set bound can be written as follows. Given a cut $\Omega \subseteq V$, let $D_{\Omega}$ be the set of demands separated by the cut, i.e., $D_{\Omega} := \{k: s_k \in \Omega, t_k \in \Omega^c\}$.
The cut-set bound bounds the sum of rates of sources in $D_{\Omega}$ and can be written as \footnote{While there is a stronger way of writing this bound, this weaker form of the bound will suffice for the purposes here.}
\beqa \sum_{k \in D_{\Omega}} R_k \leq \sup_{p_{x_1,...,x_n}} \text{I}(X_{\Omega};Y_{\Omega^c} | X_{\Omega^c}). \eeqa

Our focus on cut-set bounds as opposed to specialized  outer bounds for specific wireless channels (such as the broadcast and interference channels) is   motivated due to the following reasons.
\bit \item {\em Generality}: The cut-set bound \cite{Cuts} is an information theoretic outer bound on the achievable rate region and it can be written down for a general wireless network.
\item {\em Decomposition}: The chain rule of mutual information allows the cut-set bound of a network to decompose into the cut-set bounds on local channels; thus solutions that come close to the cut-set bound at a local level have a potential to be layered and be still close to the cut-set bound at a global level.  Formally, if we have a cut $\Omega \subseteq V$, the value of the cut is given by
\beqa \text{I}(X_{\Omega};Y_{\Omega^c} | X_{\Omega^c}) & = & H(Y_{\Omega^c} | X_{\Omega^c}) - H(Y_{\Omega^c} | X_{V})\\
& = & H(Y_{\Omega^c} | X_{\Omega^c}) - \sum_c H(Y^c_{\Omega^c} | X^c_{V})\\
& \leq & \sum_c H(Y^c_{\Omega^c} | X_{\Omega^c}) - H(Y_{\Omega^c} | X_{V})\\
& \leq & \sum_c \text{H}(Y^c_{\Omega^c} | X^c_{\Omega^c}) - \text{H}(Y_{\Omega^c} | X_{V}) \\
& = & \sum_c \text{H}(Y^c_{\Omega^c} | X^c_{\Omega^c}) - \text{H}(Y_{\Omega^c} | X^c_{V}) \\
& = & \sum_c \text{I}(X^c_{\Omega};Y^c_{\Omega^c} | X^c_{\Omega^c}), \label{eq:cut_breaks} \eeqa
and thus the cut-set decomposes into sum of the cut-sets evaluated for each channel.
     \item {\em Structure}: Cut-set bounds have been well studied in the theoretical computer science literature and their combinatorial structure have been well understood. In fact, algorithms for approximately computing the cut-set bounds form an integral part of the theory of approximation algorithms.
          \item {\em Invariance under feedback}: The cut-set bound (evaluated under general joint distributions) is essentially a bound based on upper bounding the rate by the rate of a point-to-point channels and is, therefore, invariant to feedback.
           \eit
Finally,  {\em reciprocity} of the local channels (rate region reciprocity with the roles of transmitter and receiver reversed)  will be paid attention to. The combinatorial structure imposed by the bidirected nature of each local channel will yield to efficient algorithms that are close to cuts.

\subsection{Layering Methodology}

 Layering architectures stitch together the local solutions into a global solution:
 \ben
\item The solution to a local channel allows for reliable digital
  communication at a local
  level. 
 \item Replacing each local channel by a set of (wireline) links leads
   to a network comprised of noiseless channels, with the rates on the
   various edges are {\em coupled} by the rate regions of the local
   solutions. Reciprocal local solutions ensure that the network
   obtained is {\em bidirected} (i.e., any edge between node a to
   node b has a corresponding edge between node b to a, with the two
   edges being involved in the same types of capacity region
   constraints).
 \item Over the resulting wireline network, we might have to
   potentially employ network coding to re-encode the information
   between local channels. We utilize the combinatorial properties of the coupled rate constraints to study these new class of wireline networks. In particular, if the combinatorial structure governing the rate constraints is a specific form of a polytope known as a {\em polymatroid}, we obtain polymatroidal networks.  Therefore, we study polymatroidal networks (which have local polymatroidal constraints on rate region) and prove that routing can achieve the cut-set bound to within a $\bigO (\log k)$ factor for the $k$-unicast problem, and also prove some better approximations for more specific communication problems.
 \item Since the cut-set bound on a network of channels decomposes
   into a {\em sum} of cut-set bounds on the local channels, we can
   readily compare the performance of the layering architecture to a
   fundamental upper bound on the global network performance.  \een

   Whenever local solutions are close to the cut-set bounds for the
   corresponding local channels, we can establish the fundamental
   near-optimality of the layering architecture. We have accomplished
   this program for several canonical local wireless channels including
   broadcast erasure channels, Gaussian uplink and downlink channels,
   and interference channels with diversity (example: fast fading).

\section{Polymatroidal Networks \label{sec:poly}}

Polymatroidal capacity networks and max-flow min-cut results for polymatroidal networks form the back-bone of this paper. In this section, we give a brief introduction to polymatroidal networks. A detailed discussion of polymatroidal networks and the max-flow min-cut approximation result can be found in \cite{CKRV}.

\subsection{Polymatroids}
A set function $f:2^N \rightarrow \mathbb{R}$   over a finite ground set $N$ is submodular iff $f(A) + f(B) \ge f(A
  \cap B) + f(A \cup B)$ for all $A, B \subseteq N$; equivalently
  $f(A\cup \{i\}) -f(A) \ge f(B \cup \{i\}) - f(B)$ for all $A \subseteq
  B$ and $i \not \in A$. It is monotone if $f(A) \le f(B)$ for all $A
  \subset B$. A polymatroid refers to the following set in $\mathbb{R}^N$:
  \beqa \mc{P} = \lbr (x_1,...,x_N): \sum_{i \in S} x_i \leq f(S) \quad \forall S \subseteq [N] \rbr, \eeqa
  where $f(S)$ is a monotone submodular function with $f(\emptyset) = 0$. Thus a polymatroid is fully specified by specifying a monotone submodular function with $f(\emptyset) = 0$ (we will call such a function itself as a polymatroid). An example of a polymatroid is the following: given a set of $N$ vectors $v_1,...,v_N$ in $\mathbb{R}^M$, the function $f(S)$ defined as the rank of the matrix composed of $\{v_i\}_{i \in S}$ defines a polymatroid (we refer the reader to \cite{Oxley} for an introduction on polymatroids) .

\subsection{Definition of Polymatroidal Networks}
A  commonly studied wireline scenario is one where each edge is labeled by a capacity: this is the largest amount of information flowing on that edge. Here we are interested in  a more general model which is able to handle the additional constraints when edges meet at a node, similar in spirit to the broadcast and superposition constraints in wireless.

Consider a node $v$ in a directed graph $G$ and let $\delta^-_G(v)$ be the set of edges in to
$v$ and $\delta^+_G(v)$ be the set of edges out of $v$.  In the
standard model each edge $(u,v)$ has a non-negative capacity $c(u,v)$
that is independent of other edges.  In the polymatroidal network for
each node $v$ there are two associated submodular functions:   $\rho^-_v$ and $\rho^+_v$ which impose joint capacity constraints on the edges in $\delta^{-}_G(v)$ and $\delta^+_G(v)$ respectively. That is, for any
set of edges $S \subseteq \delta^-_G(v)$, the total capacity available
on the edges in $S$ is constrained to be at most $\rho^-_v(S)$,
similarly for $\delta^+_G(v)$. Note that an edge $(u,v)$ is influenced
by $\rho^+_u$ and $\rho^-_v$.

A cut $\Omega \subseteq V$ partitions the set of vertices into two sets $\Omega$ and $\Omega^c$. Let $F(\Omega)$ denote the set of edges going from $\Omega$ to $\Omega^c$ (we will drop the dependence on $\Omega$ if it obvious from the context). The cost/capacity of $F$ is complex to
define in polymatroidal networks (note that in standard networks it is
simply $\sum_{e \in F} c(e)$ where $c(e)$ is the capacity of $e$). To
define the cost in polymatroidal networks, each edge $(u,v)$ in $F$ is
first assigned to either $u$ or $v$; we say that an assignment of
edges to nodes $g: F \rightarrow V$ is {\em valid} if it satisfies
this restriction.  A valid assignment partitions $F$ into sets $\{
g^{-1}(v) \mid v \in V\}$ where $g^{-1}(v)$ (the pre-image of $v$) is
the set of edges in $F$ assigned to $v$ by $g$. Then we minimize over
all assignments.

\begin{defn} {\em Cost of edge cut:} Given a directed polymatroidal
  network $G=(V,E)$ and a set of edges $F \subseteq E$, its cost
  denoted by $\nu(F)$ is $\min_{g: F \rightarrow V, \text{~$g$ valid}}
  \sum_v \left(\rho_v^-(\delta^-(v) \cap g^{-1}(v)) +
    \rho^+_v(\delta^+(v) \cap g^{-1}(v))\right)$.  In an undirected
  polymatroid network $\nu(F)$ is $\min_{g: F \rightarrow V,
    \text{~$g$ valid}} \sum_v \rho_v(g^{-1}(v))$.
\end{defn}

 A  max-flow min-cut theorem for the unicast problem in directed polymatroidal networks is known in the literature \cite{LawlerMartel} . This result has been extended to the broadcast traffic scenario in \cite{polymatroidBC}. Our focus here is on multiple unicast traffic, for which flow-cut gaps are currently unknown. The best known result for multiple-unicast, even in the traditional wireline network with capacity constrained edges, is the approximate optimality of max flow (in terms of being close to the min  cut); this result is available (in a strong sense) only for bidirected (or undirected) networks. Our goal is to obtain a similar result for bidirected polymatroidal networks.

 We define  a bidirected polymatroidal network as a directed polymatroidal network with the following properties.
\beit \item Every edge $e = (i,j)$ has a corresponding reverse edge $\tau(e) := (j,i) $.
\item For any vertex $v$, the polymatroidal constraint $\rvin(\cdot)$ on the incoming edges $\In(v)$ is the same as the polymatroidal constraint $\rvout(\cdot)$ on the outgoing edges $\Out(v)$. More concretely,
\beqa \rvin(E_v) = \rvout(\tau(E_v)) \quad \forall E_v \subseteq \In(v). \eeqa
\eeit

\subsection{Main Result}
The following theorem is proved in \cite{CKRV}, which generalizes the results of \cite{LeightonRao} to the case of polymatroidal capacity networks:
\bthm \label{thm:bidirected} For a bidirected polymatroidal network with $k$ source-destination pairs, the ratio between the max-flow rate region  and the min-cut rate region is $\bigO ( \log k)$. The max-flow and an approximate min-cut can be calculated in polynomial time. Furthermore, this factor is tight in general, i.e., there are families of polymatroidal networks such that the flow-cut gap is $\Omega(\log k)$. \ethm

\bpf (High level idea)
The proof is done for undirected polymatroidal networks, whose capacity is within a factor $2$ of bidirected capacity networks. The max-flow can be written down as a large linear program. While this program has an exponential number of constraints, a polynomial time algorithm can be obtained to solve this program using the polymatroidal structure. The dual of the linear program is related to the cut via an integer relaxation.  An important step in the proof is to simplify the constraint structure of the dual via the use of continuous extensions
of submodular functions, in particular the Lov\'asz extension \cite{Lovasz83}.  The resulting program has a convex objective function but the constraint structure is much simpler. This convex program (which equals the dual of the flow) and the cut are related by an integer relaxation.

The work of Linial, London and Rabinovich \cite{LLR} showed a fundamental (and {\em tight}) connection
between flow-cut gaps in edge-capacitated undirected networks and
low distortion embeddings\footnote{An embedding of a
  metric space $(V,d)$ (metric $d$ on points $V$) into another metric
  space $(V',d')$ is a mapping $f:V\rightarrow V'$. The distortion of
  $f$ is the smallest $\alpha \ge 1$ such that $\frac{1}{\alpha} \cdot
  d(u,v) \le d'(f(u),f(v)) \le \alpha \cdot d(u,v)$ for all $u,v \in
  V$.}  of finite metric spaces into $\ell_1^m$.
This connection effectively reduced the flow-cut gap question to investigating $\ell_1^m$ embeddings of finite metric spaces induced by graphs.

For the polymatroidal network problem, embeddings into a general $\ell_1$ spaces are no longer sufficient, and we require a specific type of embedding called line embedding, where the nodes are embedded into a line in $\mathbb{R}$. These embeddings were implicit in the seminal work of Bourgain \cite{Bourgain}, first explored by Matousek and Rabinovich \cite{MatousekR01} \cite{Rabinovich03} and later exploited in \cite{FeigeHL06} to obtain flow-cut gaps in undirected node-capacitated graphs. These results imply that every finite metric space of $n$ elements can be embedded by a contraction into a line preserving the average distortion of $k$ points to within a factor of $\bigO(\log k)$. We use this result to connect the convex program written using the Lov\'asz extension to the cut, hence proving a flow-cut gap of $\bigO(\log k)$.

The converse part, i.e., the existence of families of polymatroidal networks with a gap of $\Omega (\log k)$ follows from the existence of families of {\em edge capacitated} networks (which are special cases of polymatroidal networks) with a flow-cut gap of $\Omega (\log k)$.

\epf

\subsection{Special Traffic Scenarios \label{sec:specialTraffic}}

While in general, the factor of $\bigO (\log k)$ for flow-cut gaps is tight for multiple-unicast in bidirected polymatroidal networks, there may be special communication scenarios for multiple unicast when the factor can be improved. We present some instances here, where the flow cut gap is much better even for the more general case of {\em directed} polymatroidal networks.

\subsubsection{Broadcast traffic:}  Broadcast traffic is a special type of multiple unicast traffic where all the messages originate at a single source. Consider a directed polymatroidal network with a single source $s$ having independent messages to $K$ destination nodes $t_1,...,t_K$.

\blem \label{lem:directedBC} \cite{polymatroidBC} For a directed polymatroidal network with broadcast-traffic pattern, the rate region of the max flow equals the rate region of the min-cut.  \elem

\subsubsection{Sum rate in directed $X$ networks:} Consider $J$ sources $S_1$,...,$S_J$ and $K$ destinations $T_1$,...,$T_{K}$, where each source has an independent message for each destination. The rate tuple is a $JK$ length vector $R_{jk}$ between each $j$ and $k$. This communication problem is referred to commonly as the $X$-network problem.

\blem \label{lem:directedX}  For a directed polymatroidal network with $X$-traffic pattern, the {\em sum-rate} of max flow equals the sum-rate bound given by min-cut.  \elem

\bpf Construct a super source $S$ which talks to the $J$ sources with infinite capacity links and a super sink $T$ which is connected from each of the $K$ sinks via infinite capacity links.  The max-flow min-cut theorem for unicast between $S$ and $T$ in directed polymatroidal networks \cite{LawlerMartel} implies the desired result. \epf

\subsubsection{Sum rate for group communication in directed networks:} Consider a directed polymatroidal network with a specially marked group of nodes $S \subseteq V$. Each node $s$ in $S$ has an independent message for every other node in $S$. Thus it is a multiple unicast problem with $|S|(|S|-1)$ messages. We refer to this traffic pattern as the group-communication traffic pattern. Suppose we are interested only in maximizing the sum-rate.

\blem \label{lem:directedGroup} \cite{CKRV}  For a directed polymatroidal network with a group-communication traffic pattern, the sum-rate of max-flow is greater than {\em half} the sum-rate bound given by min-cut.  \elem

\bpf The proof of this theorem is non-trivial and requires a reduction from the directed polymatroidal network to the directed edge capacitated network using a combinatorial uncrossing argument. For a directed edge-capacitated network, this theorem is proved by Naor and Zosin \cite{NaorZ01}. For a detailed proof of this statement for polymatroidal networks, we refer the reader to \cite{CKRV}. \epf

\section{Local Physical Layer Schemes \label{sec:local_schemes}}

In this section, good local physical layer schemes for several channel models will be discussed. For each of these channels, the goals will be to identify a physical layer scheme, quantify its rate region, understand its closeness to the cut-set bound and to examine its combinatorial structure. We will also analyze if the rate region remains (approximately) the same when the sources and destinations are exchanged and channels are reversed. In later sections, these properties will allow us to stitch together local physical layer schemes to get global schemes. The results in this section for various channel models are summarized in Table~\ref{table:sofa}.

We will use the notation $\mc{R}^{\text{ch}}_{\ach}$ and $\mc{R}^{\text{ch}}_{\cut}$ to denote the achievable and the cut-set {\em rate regions} respectively, where $\text{ch}$ denotes the channel of interest.

\begin{table*}
\begin{center}
\begin{tabular}{||c|c|c|c||}
\hline \hline
Characteristic /      & Closeness-to-Cut & Combinatorial & Reciprocity\\
Channel         & & Structure & \\
\hline \hline
&&&\\
Linear Deterministic MAC / BC & Exact& Polymatroidal & Exact\\
\hline
&&&\\
Gaussian MAC / BC & Approximate & Polymatroidal & Approximate\\
\hline
&&&\\
Erasure Broadcast & Far & Polytope & Far \\
\hline
&&&\\
Erasure Broadcast (Feedback)  & Approximate & Polymatroidal & Approximate\\
\hline
&&&\\
Fading MAC / BC &&&\\
with delayed CSI & Approximate & Polymatroidal & Approximate\\
\hline
&&&\\
Fading linear deterministic & Approximate & Polymatroidal & Approximate \\
\hline
&&&\\
Fading X-Channel & Approximate & Polymatroidal & Approximate \\
\hline
&&&\\
Fixed X-Channel & Approximate & Polymatroidal & Approximate\\
& in DOF & DOF region & in DOF\\
\hline
\end{tabular}
\end{center}
\caption{Canonical wireless channels, as viewed via three lenses.}
\label{table:sofa}
\end{table*}

\subsection{Linear Deterministic Broadcast and Multiple Access Channels \label{sec:ld_duality}}
Consider a broadcast channel with $d$ receivers of the form, \beqa y_{i} = H_{i} x \ , \quad \forall i = 1,2,...,d, \eeqa where $y_i$ is the received vector at receiver $i$ and $x$ is the transmitted vector.
The source intends to communicate independent messages to each of its destinations. For a subset $K$ of $\{1,2,...,d\}$, let $H_K$ denote the matrix with $H_i, i \in K$ stacked up alongside one another. The capacity region of this broadcast channel \cite{Marton} is given by
\beqa \mc{C} = \{ (R_1,...,R_d): \sum_{i \in K} R_i \leq \text{Rank}(H_K) \quad \forall K \subseteq [d] \}. \label{eq:ld_bc} \eeqa
This capacity region is also equal to the cut-set bound, which is a polymatroid (see \cite{Oxley}).

Let us consider a ``reciprocal'' multiple access channel in which there are $d$ transmitters and one receiver,
 \beqa y = \sum_{i} H_{i}^T x_i, \eeqa
  and all the transmitters have an independent message to transmit to the single destination. The capacity region of a general MAC channel is known (see, for example, Chapter 14 in \cite{CovTho}) and for the linear deterministic channel, is given again by the rate region in \eqref{eq:ld_bc}. We observe that the capacity of the broadcast channel and the reciprocal MAC channel are the same and are equal to their cut-set bound.

Thus for a linear deterministic broadcast and MAC channel, the rate region is exactly polymatroidal, equal to the cut, and is reciprocal.

\subsection{Gaussian Broadcast and Multiple Access Channels \label{sec:g_duality}}

Let us first consider a multiple access channel, defined by
 \beqa y = \sum_{i} h_{i} x_i + z, \eeqa
  where the transmitted vector $x_i$ is constrained by a power constraint $P$ at each of the $d$ nodes, $y$ is the received vector and $z$ denotes the noise, which is of unit power. Let the rate region achievable on this multiple access channel be denoted by $\mc{R}^{\text{MAC}}_{\text{ach}}(P)$. This region is known to be polymatroidal \cite{TH}. This rate region equals the cut-set bound evaluated under product distributions \cite{CovTho} , i.e., \beqa \mc{R}^{\text{MAC}}_{\text{ach}}(P) =  \mc{R}^{\text{MAC}}_{\text{cut,product}}(P) \label{eq:mac_rate_cut}. \eeqa Let the cut-set bound evaluated under general distributions be given by $\mc{R}^{\text{MAC}}_{\text{cut,general}}(P)$. We can easily verify the relation,
  \beqa \mc{R}^{\text{MAC}}_{\text{ach}}(P) \supseteq  \mc{R}^{\text{MAC}}_{\text{cut,general}} \lpr \frac{P}{d} \rpr. \eeqa

Next, let us consider a ``reciprocal'' or ``dual'' broadcast channel, given by,
\beqa y_{i} = H_{i} x + z_i \ , \quad \forall i = 1,2,...,d, \eeqa
where the transmitted vector $x$ is constrained by a power constraint $dP$, $y_i$ is the received vector and  $z_i$ denotes the noise at each receiver, which is of unit power. Let us call the rate region of the broadcast channel as $\mc{R}^{\text{BC}}(P)$. This rate region has been fully characterized, but is not equal to the cut-set bound and is not polymatroidal (see Chapter $6$ in \cite{TseVis} for a discussion).

The rate region of this broadcast channel with sum power constraint $P d$ contains the rate region of the multiple access channel with {\em individual constraint} $P$ at each node \cite{VT03,VJG03}, i.e.,
\beqa  \mc{R}^{\text{BC}}_{\ach}(P) \supseteq \mc{R}^{\text{ MAC}}_{\ach}(P). \eeqa
For purpose of symmetry, we can choose to operate the broadcast channel at the rate region specified by $\mc{R}^{\text{MAC}}(P)$, i.e., let us set \beqa \mc{R}^{\text{BC}}_{\text{ach}}(P) = \mc{R}^{\text{MAC}}_{\text{ach}}(P) \label{eq:bcmac_rate}. \eeqa This will also ensure that the rate region of the achievable scheme $\mc{R}^{\text{BC}}_{\text{ach}}(P)$ is polymatroidal. Thus, in our achievable strategy, the rate region of a multiple access and that of the dual broadcast channel are equal and given by a polymatroidal region.

Let the cutset bound for the broadcast channel be specified as $\mc{R}^{\text{BC}}_{\text{cut,general}}(P)$. Since there is only one input variable, the cut-set bound under product distribution and general distribution are the same.
\blem \label{lem:cuts_poly_gauss} The achievable region of the MAC channel compares with the cutset bound under general distributions as follows,  \beqa \mc{R}^{\text{MAC}}_{\text{ach}}(P) \supseteq  \mc{R}^{\text{MAC}}_{\text{cut,general}} \lpr \frac{P}{d} \rpr. \eeqa
For the broadcast channel, we have the relation,
\beqa \mc{R}^{\text{BC}}_{\text{ach}}(P) \supseteq \mc{R}^{\text{BC}}_{\text{cut,general}} \lpr \frac{P}{d} \rpr. \eeqa \elem

\bpf The proof is deferred to Appendix \ref{app:cuts_poly_gauss}. \epf

Thus for a Gaussian broadcast and MAC channel, the rate region is approximately polymatroidal, close to the cut, and is approximately reciprocal.

\subsection{Broadcast Erasure Channels \label{sec:BEC}}

Consider a network comprised of broadcast erasure channels. For a broadcast erasure channel with $d$ receivers, the channel model can be written as,
\beqa y_i = e_i x, \quad \forall i = 1,2,...,d, \eeqa
where $e_i$ is a binary random variable which when $0$ represents that at receiver $i$, the packet got erased.
If $e_i$ are all independent, then the broadcast erasure channel is said to be an independent erasure broadcast channel. Such a channel is specified by erasure probabilities $\epsilon_i,i=1,2,...,d$, where \beqa \epsilon_i = \text{Pr} \{ e_i = 0\}. \eeqa For the purpose of simplicity in this paper we will consider only broadcast erasure channels that are independent and symmetric, which implies that there is only one parameter $\epsilon$ and $\epsilon_i = \epsilon \quad \forall i$.

\blem For an erasure broadcast channel without feedback, the cut-set bound can be as large as $L$ times the achievable rate region. \label{lem:BEC_No_FB} \elem

\bpf See Appendix~\ref{app:BEC_No_FB}. \epf

This result implies that for broadcast erasure channels (without feedback), there are no good local schemes that can achieve close to the cut.

\subsubsection{Broadcast Erasure Channels with Feedback}

Since there are no good local schemes for the broadcast channel, we suggest that the scale of the locality be enlarged to include the presence of feedback  links in the physical model to get better local schemes.

The capacity of the erasure broadcast channel with ACK feedback (the receiver acknowledges whether it received the packet) was studied by \cite{Tas2User} for the two user channel, and later extended to the more general  case independently by \cite{Wang} and \cite{TasKUser}. The schemes are based on network coding and interference alignment and demonstrate that the following rate region is achievable.
\blem \label{lem:BEC_FB} \cite{Wang,TasKUser} The following rate region is achievable for the erasure broadcast channel with ACK feedback,
\beqa R_{\text{ach,fb}} = \lbr (R_1,...,R_D) | \sum_{i =1,2,..,d} \frac{R_{\pi(i)}}{1- \epsilon^i} \leq 1 \quad \forall  \pi \rbr. \eeqa \elem
Here $\pi$ is a permutation of the  set $(1,\ldots ,d)$.
Note that this region is not polymatroidal. However, it is close to the min cut rate region (which is itself polymatroidal) as seen below:
\blem \label{lem:BEC_FB_Cut}  \beqa \mc{R}_{\text{ach,fb}} \supseteq \frac{ \mc{R}_{\text{cut}}}{\bigO ( \log d)}. \eeqa \elem
\bpf See Appendix~\ref{app:BEC_FB_Cut}. \epf

The reciprocal nature of wireless channels from which the broadcast erasure channel is constructed naturally suggests a way of providing feedback links of {\em commensurate} strength. Formally: a channel is said to have {\em commensurate feedback} if there is are feedback links from the various receiving nodes to the transmitters with the same rate region as the cut-set bound for the forward channel. In Appendix~\ref{app:MAC_Erasure}, we look at one possible way of obtaining feedback links of commensurate strength as the forward links.

Thus for a broadcast erasure channel with feedback, the rate region is approximately polymatroidal, close to the cut, and is approximately reciprocal.

\subsection{MIMO Broadcast and MAC Channel with Delayed Feedback \label{sec:Delayed_BC_MAC}}

We will consider MIMO broadcast and MAC channels, similar to Sec.~\ref{sec:g_duality}, but with the difference that the channel states are i.i.d. over antennas and time. If the CSI is instantaneously available at the transmitter and the receiver, then the methods and results of Sec.~\ref{sec:g_duality} continue to hold. However, the problem becomes interesting when global channel state information is no longer available. In some settings, the channel may change so fast that by the time, the channel state feedback reaches the transmitter, the channel state has changed significantly. This feature is essentially present in the erasure channel if we think of erasure as being a channel state, which is unknown to the transmitter originally but the presence of the ACK feedback delivers this channel state to the transmitter with a delay. Recent work by Ali and Tse \cite{AliTse} showed the surprising result that even in Gaussian networks, delayed CSI can be very beneficial as compared to the absence of CSIT \cite{YatesTse}. This will form the basis of our investigation of these channels.

We assume that each broadcast and MAC channel gets feedback from its receivers about the channel state, however, the channel changes before the feedback arrives, precluding the use of feedback to predict the future state of the channel. We will resort to a degree-of-freedom characterization for this problem. Let $d_i$ denote the achievable degrees of freedom for the $i$-th message, i.e.,
\beqa d_i := \lim_{\SNR \rightarrow \infty}  \frac{R_i(\SNR)}{ \log \SNR }, \eeqa
where $R_i$ is the achievable rate for user $i$. The achievable DOF region is denoted by $\mc{D}_{\text{ach}}$, and the region given by the cutset bound is denoted by $\mc{D}_{\text{cut}}$.

Such a multiple access case is easy to deal with because even without channel state information, the cut-set bound can be achieved,
 \beqa \mc{D}^{\text{MAC}}_{\text{ach}} = \mc{D}^{\text{MAC}}_{\text{cut}}. \eeqa
 However for the corresponding broadcast channel, in the absence of CSIT, we can only achieve rates that are far from the cut-set. In \cite{AliTse}, it is shown that the presence of channel state feedback, even when delayed, can significantly alter the situation, as was the case with broadcast erasure channels:
\blem \label{lem:AliTse} \cite{AliTse} For a fading MISO broadcast channel with a source with $L$ transmit antennas and $L$ single transmit antenna receivers, the following DOF region can be achieved with the help of delayed CSIT:
\beqa \mc{D}^{\text{BC}}_{\ach} = \left\{ \sum_{i=1}^L D_i  \leq l \frac{K}{\sum_{i=1}^L \frac{1}{i}} \right\}. \eeqa  \elem

Our first result is to obtain an approximation for the rate region of MIMO broadcast channels with delayed feedback, formally stated in the following lemma.
\blem \label{lem:Fading_MIMO} For a fading MIMO broadcast channel with $l$ transmit antennas and $d$ users with user $i$ having $m_i$ antennas, and delayed CSI feedback, the following DOF region can be achieved:
\beqa \mc{D}^{\text{BC}}_{\text{ach}} \supseteq \frac{\mc{D}^{\text{BC}}_{\text{cut}}}{\bigO (\log p)}, \eeqa
where $ p := \min (l, \sum_i m_i) $ is the minimum of the number of transmit and receive antennas in the system.  \elem

Thus for a fading MIMO broadcast channel, the rate region is approximately polymatroidal, close to the cut, and is approximately reciprocal.

\subsection{Fading $X$-channels  \label{sec:singleHop}}
Consider a $L$-user $X$-channel, there are $L$ sources $s_1,...,s_L$ and $L$ destinations $t_1,...,t_L$ with {\em each} source having message to send to each destination, thus there are $L^2$ messages in total. The connectivity graph between the nodes is a bi-partite graph with $E$ being the set of edges. We will abbreviate edge $(s_i,t_j)$ as $(i,j)$ since the meaning is clear from the context.

Note that an interference channel is a special case of this channel. Capacity achieving schemes even for the $L$-user interference channel are not known in the general setting. In  \cite{CadJafIA}, the authors show that each user in a $L$-user interference channel can achieve half their point-to-point degrees-of-freedom (DOF) if the channel is fast-fading, using a mechanism called interference alignment, which was initially proposed for the $2$-user $X$-channel in \cite{MMK}. This result has since been generalized in several directions; most notably, in \cite{ErgodicIA}, it is shown that using an ``ergodic interference alignment'' scheme, each user can get half her rate at {\em all SNR}, and in \cite{RealIA1}, it is shown that the DOF result can be proved even under {\em fixed channel coefficients} using a scheme termed ``real interference alignment''. These results have been unified into a single framework in \cite{WuVerdu}. It has also been shown \cite{NiesenScaling} that ergodic interference alignment can be used to achieve linear capacity scaling in dense interference networks.

\subsubsection{Cuts in an $X$-channel}

\begin{figure}[htb]
\begin{centering}
\scalebox{0.6}{\includegraphics{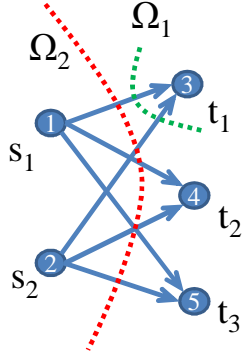}}
\caption{Cuts in $X$-Channel} \label{fig:cuts}
\end{centering}
\end{figure}

Consider an example $X$-channel with two sources and three destinations, shown in Fig.~\ref{fig:cuts}. Two cuts are marked in the figure. The light (green) cut $\Omega_1$ separates only destination $1$ from the two sources, thus providing a bound: $R_{11}+R_{21} \leq C_{\Omega_1}$, whereas dark (red) cut $\Omega_2$ separates all sources from all destinations and therefore provides a bound on $R_{11}+R_{12} +R_{13} +R_{21}+R_{22} + R_{23} \leq C_{\Omega_2}$. Of these two bounds, the first bound corresponds to that of a polymatroidal constraint whereas the second bound does not correspond to a polymatroidal constraint (since in a polymatroidal network, only edges that meet at a node have a joint constraint).

For a general $X$-channel, these two types of cuts will be present, and we can classify them as
\ben \item Cuts that separate a single node from the rest of the nodes (referred to as cuts of the polymatroidal form), and
 \item Cuts that separate multiple sources from multiple destinations.
 \een

We would like to show an achievable scheme that not only achieves the cut-set bound approximately, but also the rate region of the achievable scheme satisfies a polymatroidal constraint. Therefore, for an $X$-channel, we will have to show that only cuts of the polymatroidal form (separating one node from the rest) play a dominant role. This is a key challenge that we address in this section.

\subsubsection{Channel Model \label{sec:channel_model}}
The channel model can therefore be written as,
\beqa y_i(t) = \sum_{j \in \In(i)} h_{ij}(t) x_j(t) + z_i(t) \quad \forall t=1,2,...,T, \eeqa
where $x_i(t),y_i(t),z_i(t)$ are the transmitted vector, received vector and noise vector at time $t$,  $\In(i)$ represents the set of neighbors of node $i$ who have an incoming edge to $i$ and fading coefficient $h_{ij}(t)$ is associated with  edge $(i,j) \in E$ at time $t$. The noise vector is assumed to have unit variance and is independent at each node. There is a power constraint of $P$ per node.

We will make the following assumptions about the fading distribution:
 \beit \item Fading coefficients are assumed to be i.i.d. over edges and over time.
 \item The fading coefficient will be assumed to be symmetric, i.e., if $h_{ij}$ is a discrete random variable, \beqa \Pr \{h_{ij} = a \} = \Pr \{h_{ij} = -a \}, \quad \forall a \eeqa otherwise, if the random variable is absolutely continuous, the pdf $p(.)$ must satisfy \beqa p (h_{ij} = a) = p( h_{ij} = -a ) \quad \forall a . \eeqa
\item The fading distribution is assumed to satisfy:
\beqa a := e^{-\mb{E}( \log  |h|^2)} < \infty . \eeqa
\eeit

One example of a fading distribution that satisfies these assumptions is when $h_{ij}(t)$ is i.i.d. across nodes and time with a complex gaussian distribution, for which $a = 1.723$ \cite{Oyman}.

 We will use the shorthand $C(P)$ to denote the ergodic capacity of a fading channel with power constraint $P$, and the fading coefficient $h$ of unit variance,
\beqa C(P) :=  \mb{E}_{h} \left [ \frac{1}{2} \log ( 1 + |h|^2 P) \right ]. \eeqa

\subsubsection{Scheme for the $K$-user interference channel} First, we consider the case of $k$-user interference channel, where there are messages only from $s_i$ to $t_i$ for each $i=1,2,...,k$. In the ergodic set-up, the following result is known:
\blem \label{lem:ergodicIA} \cite{ErgodicIA} For an $k$-user ergodic interference channel, where the direct links are non-zero, the following rate tuple is achievable:
\beqa R_i = \frac{1}{2} C(2 P), \quad \forall i=1,2,...,k. \eeqa \elem

\subsubsection{Scheme for the $X$-channel} We generalize this physical layer scheme to the $X$-channel with $L$ sources and $M$ destinations, demonstrating not only that the cut-set bound is approximately achievable, but also that only cut of the polymatroidal form are relevant.

\bthm \label{thm:single_layer_X} For an $L$-source, $M$-sink ergodic $X$-channel, the following rate region is achievable,
\beqa \mc{R}_{\text{X}}^{\ach} = \left\{ (R_{ij}) |
\begin{array}{c}     \sum_{j : (i,j) \in E} R_{ij} \leq \frac{1}{2} C(2 P) \quad \forall i \in S \\
                     \sum_{i : (i,j) \in E} R_{ij} \leq \frac{1}{2} C(2 P) \quad \forall j \in T \end{array} \right\}. \label{eq:single_layer_X}      \eeqa
and furthermore, if $d$ is the maximum degree of any node,
\beqa \mc{R}^{\text{X}}_{\ach} (P) \supseteq \frac{\mc{R}^{\text{X}}_{\cut}( \frac{2P}{ad})}{2}, \eeqa where $a:=e^{-\mb{E}( \log  |h|^2)}$. \ethm

\bpf Let us write $R_{ij}$ for the rate of communication between $s_i$ and $t_j$. We use the following achievable strategy:

\beit

\item Let us construct a bipartite graph between the source vertices and sink vertices, and edges given by $E$.
\item A matching in a bipartite graph is a choice of edges such that each node is present in at most one edge. In our case, a matching can be thought of representing a choice of at most one destination for each source.
     Choose a matching $M$ on the bipartite graph, let $\pi$ be the corresponding permutation. The characteristic vector of a bipartite matching is given by the vector $(x_{ij}): x_{ij} = 1$, if $(i,j) \in M$, otherwise $x_{ij} = 0$.
\item Consider the interference channel from $s_1,...,s_L$ to $d_{\pi(1)},...,d_{\pi(L)}$. For the $s_i,t_{\pi(i))}$ pairs that are connected, we can achieve a rate of $\frac{1}{2} C(2 P)$ using the strategy of Lemma~\ref{lem:ergodicIA}.
\item This implies that a rate given by $\frac{1}{2} C(2 P)$ times the characteristic vector of the bipartite matching is achievable.
\item Now, we can achieve any convex combination of the rates given by matchings on the graph. This is given by the following polytope, called the matching polytope,
    \beqa \mc{M} = \text{conv} \{ (x_{ij})_M | M \text{ a matching } \}. \eeqa
\item By a theorem in bi-partite graph matchings \cite{Schrijver}, this matching polytope can be alternately described as:
 \beqa \mc{P} = \left\{ (x_{ij}) | x_{ij} \geq 0 \quad \forall i,j,  \begin{array}{c}
 \sum_{j: (i,j) \in E} x_{ij} \leq 1 \quad \forall i \\ \sum_{i : (i,j) \in E} x_{ij} \leq 1 \quad \forall j \end{array} \right\}. \eeqa
\item Therefore the achievable rate region is given by \eqref{eq:single_layer_X}.
\item The cut-set bound $\mc{R}_{\text{X}}^{\cut}$ implies the following, which are only a subset of the cuts (the cuts which separate one node from all the others) :
     \beqa \left\{ (R_{ij}) |  \begin{array}{c}
     \sum_{j: (i,j) \in E} R_{ij} \leq  \mb{E} \log ( 1 + \sum_{j: (i,j) \in E} h_{ij}^2 P ) \quad \forall i \in S \\
     \sum_{i: (i,j) \in E} R_{ij} \leq \mb{E} \log ( 1 + \sum_{i: (i,j) \in E} h_{ij}^2 P) \quad \forall j \in T \end{array} \right\}. \nonumber \eeqa
\item

Now, due to the concavity of the logarithm,
\beqa \mb{E} \log ( 1 + \sum_{j: (i,j) \in E} h_{ij}^2 ) & \leq &     \log ( 1 + \sum_{j: (i,j) \in E} \mb{E} h_{ij}^2 P ) \\
& \leq  &  \log ( 1 +  dP) \\
& \leq & \mb{E} \log ( 1 + adP |h|^2) = C(adP)  \eeqa
where the last step follows because of the convexity of the function $f(x) = \log ( 1 + ce^{x})$, i.e., applying Jensen inequality for the aforementioned convex function, we get,
\beqa \mb{E} ( \log ( 1 + c |h|^2) ) & = &  \mb{E} \{ \log ( 1 + c e^{\log  |h|^2}) \} \\
& \geq & \log ( 1 + c e^{\mb{E}( \log  |h|^2)}) \\
& = & \log( 1+c a^{-1}),  \eeqa
where $a:= e^{-\mb{E}( \log  |h|^2)})$.

Thus the cut-set bound implies the following inequalities,
     \beqa \left\{ (R_{ij}) |  \begin{array}{c}
     \sum_{j: (i,j) \in E} R_{ij} \leq  C(adP) \quad \forall i \in S \\
     \sum_{i: (i,j) \in E} R_{ij} \leq C(adP) \quad \forall j \in T \end{array} \right\}. \nonumber \eeqa

\item Therefore we get the result that
\beqa \mc{R}^{\text{X}}_{\ach} (P) \supseteq \frac{\mc{R}^{\text{X}}_{\cut}( \frac{2P}{ad})}{2}. \eeqa
\eeit

 \epf

Thus for a fading $X$ channel, the rate region is exactly polymatroidal, approximately close to the cut, and is exactly reciprocal (since the description of the rate region remains the same even the channel is reversed).

\subsection{Fixed $X$-channels  \label{sec:fixedSingleHop}}
Consider a $L$-user $X$-channel with fixed channel coefficients drawn from a continuous distribution. We will obtain a degrees-of-freedom characterization of this $X$-channel (which holds almost surely). The channel model can therefore be written as,
\beqa y_i(t) = \sum_{j \in \In(i)} h_{ij} x_j(t) + z_i(t) \quad \forall t=1,2,...,T, \eeqa
where $x_i(t),y_i(t),z_i(t)$ are the transmitted vector, received vector and noise vector at time $t$,  $\In(i)$ represents the set of neighbors of node $i$ who have an incoming edge to $i$ and channel coefficient $h_{ij}$ associated with  edge $(i,j) \in E$ is drawn from a continuous distribution which has a probability density function, i.e., the probability measure is absolutely continuous with respect to the Borel measure. The noise vector is assumed to have unit variance and is independent at each node. There is a power constraint of $P$ per node.

First, we consider the case of $L$-user interference channel, where there are messages only from $s_i$ to $t_i$ for each $i$. The following result characterizes the degrees of freedom of the $L$-user interference channel:
\blem \label{lem:realIA} \cite{RealIA1} For a $k$-user interference channel with channel coefficients drawn from a continuous distribution, if the direct links are non-zero, the following DOF tuple is achievable almost surely:
\beqa D_i = \frac{1}{2}, \quad \forall i=1,2,...,k. \eeqa \elem

We can generalize this interference channel scheme to the $X$-channel with $L$ sources and $M$ destinations, using the same method as in Theorem~\ref{thm:single_layer_X}.

\bthm \label{thm:single_layer_X_fixed} For an $L$-source, $M$-sink fixed $X$-channel, the following DOF region is achievable {\em almost surely},
\beqa \mc{D}_{\text{X}}^{\ach} = \left\{ (D_{ij}) |
\begin{array}{c}     \sum_{j : (i,j) \in E} D_{ij} \leq \frac{1}{2}  \quad \forall i \in S \\
                     \sum_{i : (i,j) \in E} D_{ij} \leq \frac{1}{2} \quad \forall j \in T \end{array} \right\}. \label{eq:single_layer_X}      \eeqa
and furthermore,
\beqa \mc{D}^{\text{X}}_{\ach} \supseteq \frac{\mc{D}^{\text{X}}_{\cut}}{2} \quad \text{a.s.} \eeqa \ethm

\bpf The proof proceeds in a  manner quite similar to that of Theorem~\ref{thm:single_layer_X}; the only difference is that we use Lemma~\ref{lem:realIA} instead of Lemma~\ref{lem:ergodicIA}. Also, since we are dealing with DOF, which is an SNR-scaling characterization, the (constant) power scaling factor is not relevant. \epf

Thus for a fixed $X$ channel, the achievable DOF region is exactly polymatroidal, approximately close to the cut, and is exactly reciprocal (since the description of the achievable DOF region remains the same even the channel is reversed).

\subsection{Fading Linear Deterministic Channels \label{sec:fadingLD}}
In Sec.~\ref{sec:ld_duality}, we considered linear deterministic networks which had only broadcast and MAC components. In this section we will consider a general linear deterministic network with fading. The communication network is represented by a directed graph $G = (V,E)$. If $(i,j) \in E \iff (j,i) \in E$, we call the network bidirected.  The edges $(i,j)$ that are present have fading matrix $H_{ij}(t)$ on them. Each fading coefficient in each matrix is distributed i.i.d. fading over edges and time. The fading distribution for each non-zero coefficient is assumed to be uniform over the finite field (except zero) $\mb{F}_q \ {0}$. The proof can be extended to the case where the fading takes value $0$ but there will be penalty factor of $\frac{q-1}{q}$.

For a linear deterministic interference network with fading, there is a scheme based on ergodic interference alignment that achieves half the point-to-point rate for each user:

\blem \cite{ErgodicIA} For an $k$-user fading linear deterministic interference channel, with direct links being non-zero, the following rate type is achievable:
\beqa R_i = \frac{1}{2} \log_2 (q), \quad \forall i=1,2,...,k. \eeqa \elem

We can use this scheme to create a scheme for the $X$-channel, the rate region of this scheme is quantified in the following theorem.

\bthm \label{thm:LD_X} For an $L$-source, $M$-destination deterministic ergodic $X$-channel, the following rate region is achievable,
\beqa \mc{R}^{\text{X}}_{\ach} \supseteq \frac{\mc{R}^{\text{X}}_{\cut}}{2}, \eeqa
furthermore, only cuts that separate one node from the rest are sufficient. \ethm
\bpf The proof for this case is very similar to the proof of Theorem~\ref{thm:single_layer_X}, except that the term $\frac{1}{2} C( 2P)$ is now replaced by $\frac{1}{2} \log_2(q)$. Also,  there are no power scaling losses in the case of a linear deterministic network. \epf

Thus for a linear deterministic $X$ channel with fading, good local schemes exist; furthermore the schemes have a rate region which can be described by polymatroidal constraints.

\section{Approximate Capacity Results for Wireless Networks \label{sec:approx_capacity}}
In this section, we will present approximate capacity results for wireless networks with {\em multiple unicast traffic} for several channel models. Our achievable schemes will use a layering architecture and  we will use the cut-set bound as the outer bound. For each channel model, we will have the description of the wireless network as a graph $G=(V,E)$.  There are $k$ designated source nodes $s_1,...,s_k$, which have independent messages and $k$ corresponding destination nodes $t_1,...,t_k$:  $t_i$ wants to decode the message of $s_i$ with vanishingly small probability of error, for every $i=1,\ldots k$. Let $\mc{R}^{\text{ch}}_{\ach}$ denote the rate region comprising of the rate tuples achievable
and let $\mc{R}^{\text{ch}}_{\cut}$ denote the rate region corresponding to the cut-set bound for the channel model $\text{ch}$. If a power constraint $P$ is present, we will describe these regions as a function of $P$, i.e., $\mc{R}^{\text{ch}}_{\ach}(P)$ and $\mc{R}^{\text{ch}}_{\cut}(P)$.


\subsection{Gaussian networks with MAC and Broadcast components \label{sec:MacBc}}

We consider Gaussian networks in which there are only broadcast and MAC components, i.e., there is no interference channel component. This is equivalent to the assertion that  each edge is involved in either a superposition constraint or in a broadcast constraint but not in both. In practice, such a network can be realized by using a partial frequency reuse scheme, where the total bandwidth is divided into different chunks, which are assigned to users in such a way that interference component is avoided.

Gaussian networks with MAC and broadcast components alone have been previously considered in \cite{KotMedEff2}, where it has been shown that for a network composed of $2$-user MAC and broadcast channels, a separation architecture involving local coding is approximately optimal. Since this general structural result holds for all possible traffic models, it may be necessary in general to use network coding \cite{NetCod} \cite{KotMed} at the network layer. In contrast, in this paper, we will assume multiple unicast traffic and utilize the {\em reciprocity} of the wireless network to show that the cut-set bound can be approximately achieved using a separation scheme along with {\em routing}.

\hspace{0.1in}
\begin{figure}[htb]
\begin{centering}
  \subfloat[Example of Frequency planned network]{\label{fig:freqReuseNW} {\includegraphics[height=30mm]{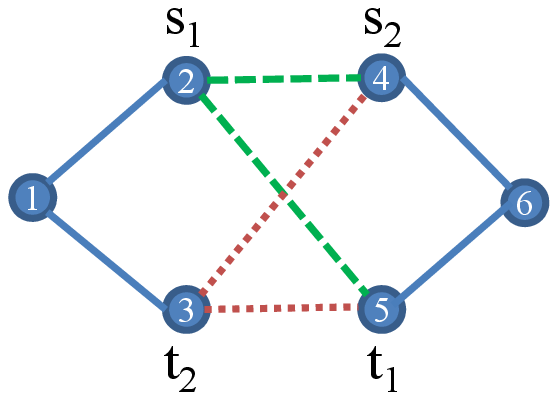}}}
 \vspace{0.2in}
 \subfloat[Equivalent noiseless network]{\label{fig:polymatroidalNW}{\includegraphics[height=40mm]{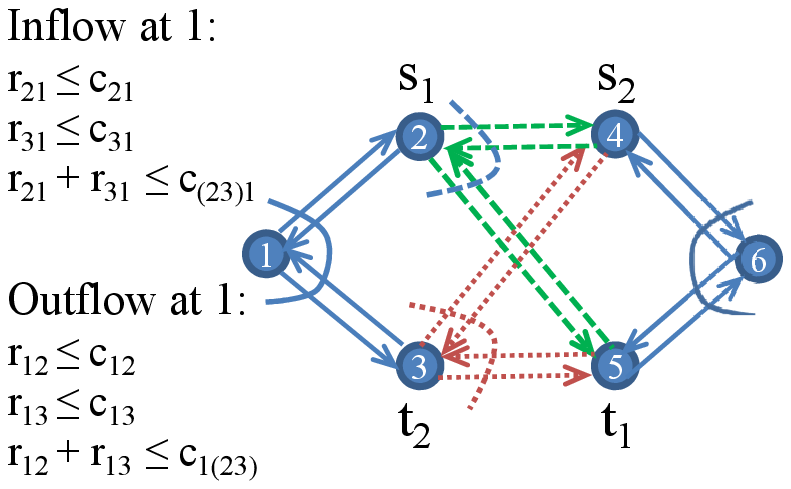}}}
\caption{A Network composed of broadcast and MAC components and its Polymatroidal Equivalent} \label{fig:coloredNW}
\end{centering}
\end{figure}

\paragraph{Network Model:} The communication network is represented by an undirected graph $G = (V,E)$, and an edge coloring $\psi: E \rightarrow C$, where $C$ is the set of colors. Each node $v$ has a set of colors $C(v) \subseteq C$ on which it transmits and receives. Each color can be thought of as an orthogonal resource  (for example, a frequency band), and therefore the broadcast and superposition constraints for the wireless channel apply only {\em within} a given color. We will assume the nodes are equipped with full duplex radios on each of these resources. For simplicity of notation, we will assume that there is a one-to-one correspondence between colors and channels, i.e., each channel operates on a distinct color; so $c$ stands for a unique color and a unique channel.

The channel model can therefore be written as,
\beqa y_i^c = \sum_{j \in \mc{N}_c(i)} h_{ij}^c x_j^c + z_i^c \quad \forall c \in C(i), \eeqa
where $x_i^c,y_i^c,z_i^c$ are the transmitted vector, received vector and noise vector on color $c$, $h_{ij}^c = h_{ji}^c$ is the channel coefficient between node $i$ and node $j$ on color $c$ and $\mc{N}_c(i)$ represents the set of neighbors of node $i$ who are operating on color $c$ and $d_c(i) = |\mc{N}_c(i)|$ be the degree of node $i$ in color $c$. Let $d = \max_{c,v} d_c(v)$ be the maximum degree of any node in a given color; therefore, $d$ is the maximum number of users on any component broadcast or multiple access channel. Each node has a power constraint $P$ {\em per edge}. Therefore node $v$ has power constraint $P d_c(v)$ for transmitting on color $c$. By the very definition, this network has a reciprocal MAC channel for every broadcast channel and vice versa. Let $V(c) = \{v: c \in C(v) \}$ be the set of nodes that use the color $c$. An example of a wireless network along with its
   equivalent noiseless network are shown in Fig.~\ref{fig:coloredNW}.

\bthm \label{thm:bc_mac} For the $k$-unicast problem in Gaussian network composed of broadcast and multiple access channels, a simple separation strategy can achieve a rate,
\beqa \mc{R}^{\g}_{\ach}(P) \supseteq \frac{\mc{R}^{\g}_{\cut}(\frac{P}{d})}{\bigO (\log k)}.  \eeqa
\ethm
This means that  the min cut, scaled down in power by a factor $d$ and in rate by a factor ${\bigO (\log k)}$, can be achieved.
 For the unicast scenario ($k=1$), we can show using a similar proof that
\beqa \mc{R}^{\g}_{\ach}(P) \supseteq {\mc{R}^{\g}_{\cut}\left (\frac{P}{d} \right )}. \eeqa
This result is similar to that obtained by \cite{ADT}, except that here it is obtained for the special case of networks composed of broadcast and multiple access channels. The scheme in \cite{ADT} requires a global physical layer scheme (the ``quantize and map" strategy), while for the special case of networks here we show that a simple separation strategy suffices.

\subsubsection{Coding Scheme: Proof of Theorem~\ref{thm:bc_mac}}
The coding scheme is a separation-based (layered) strategy: each component broadcast or multiple access channel is coded for independently creating bit-pipes on which information is routed globally. The achievable scheme used for the MAC and broadcast channel are discussed in detail in Sec~\ref{sec:g_duality}.
The achievable rate region for the MAC and Broadcast channels described there are polymatroidal and therefore
each multiple access or broadcast channel with $d$ users can be replaced by a set of $d$ bit-pipes whose rates are jointly constrained by the corresponding polymatroidal constraints. Thus we get a polymatroidal network by using this layered strategy; this polymatroidal network is described as follows: for each node $v$ in the original graph, there are several vertices $v_c$, one for each color $c \in C(v)$. There is an edge between $u_c$ and $v_c$ if $h_{uv}^c \neq 0$, the polymatroidal constraints are given by \beqa \rho_{v_c}^{-}(F_{v}) = \log (1+ \sum_{u:(u,v)\in F_v} |h_{uv}^c|^2 P) \quad  \quad \forall F_v \subseteq \delta^{-}(v_c)\\
\rho_{v_c}^{+}(F_v) = \log (1+ \sum_{u:(u,v)\in F_v} |h_{vu}^c|^2 P) \quad  \quad \forall F_v \subseteq \delta^{+}(v_c), \eeqa
and the polymatroidal network is bidirected due to the fact that $h_{uv}^c = h_{vu}^c$ and the reciprocity in the rate regions of the MAC and BC channel. Further, there are edges between $v_c$ and $v_{c'}$ of infinite capacity, since these correspond to the same node $v$ in the original graph.


Let us call the cut-set bound region on this induced polymatroidal network as $\mc{R}^{\poly}_{\text{cut}}(P)$, and let us call the rate region for the flow-based achievable scheme as
$\mc{R}^{\poly}_{\ach}(P)$. Then we have from Theorem~\ref{thm:bidirected} that
\beqa \mc{R}^{\poly}_{\text{flow}}(P) \supseteq \frac{\mc{R}^{\poly}_{\text{cut}}(P)}{\bigO \log k}. \label{eq:ind_log3} \eeqa

As an example, the bidirected polymatroidal network induced for the example of Fig.~\ref{fig:freqReuseNW} is shown in Fig.~\ref{fig:polymatroidalNW}. The submodular constraints are explicitly written down only for node $1$, but similar constraints apply at nodes $2$, $3$ and $6$. In this figure, $c_{21},c_{31}$ and $c_{(23)1}$ represent constraints on the rate of communication from node $2$ to $1$, the rate of communication from node $3$ to $1$ and the sum rate from nodes $2$ and $3$ to $1$ respectively.

It is now sufficient to compare the cuts on the polymatroidal and the Gaussian network.
\blem \beqa \mc{R}^{\text{g}}_{\text{cut}}(P)  \subseteq   \mc{R}^{\poly}_{\text{cut}}(d   P). \label{eq:cuts_sep_g} \eeqa \elem

\bpf
Given a cut $F_{\Omega}$ in the polymatroidal network, we will show that there is a corresponding cut in the Gaussian network, whose value is within a power scaling factor $d$ of the polymatroidal cut. The value of the cut in the polymatroidal network is $\nu(F_{\Omega}) = \sum_c \nu(F^c_{\Omega})$, i.e., the polymatroidal cut breaks up into the sum of the cuts of various colors. The value of cut in a given color $\nu(F^c_{\Omega})$ corresponds to a certain polymatroidal constraint in a given MAC or broadcast channel. We need to show that there is a similar cut in the Gaussian network whose value is within a power scaling factor.

%
As shown in Lemma~\ref{lem:cuts_poly_gauss}, we have that for each of these channels, the Gaussian cut and the polymatroid representing the achievable scheme are within a power scaling factor of $d$, and therefore
\beqa \mc{R}^{\text{g}}_{\text{cut}}(P)  \subseteq   \mc{R}^{\poly}_{\text{cut}}(d   P), \label{eq:cuts_sep_g} \eeqa
since both polymatroidal and Gaussian cuts decompose into sum of individual cuts.
\epf

The achievable rate using the separation strategy is given using \eqref{eq:ind_log3} as
\beqa \mc{R}^{\poly}_{\ach}(P)  \supseteq \frac{\mc{R}^{\poly}_{\text{cut}}(P)}{\bigO (\log k)} \supseteq  \frac{\mc{R}^{\text{g}}_{\text{cut}}(\frac{P}{d})}{\bigO (\log k)}. \eeqa

This completes the proof of Theorem~\ref{thm:bc_mac}.

\subsubsection{Special Traffic Scenarios}
We now present results for {\em directed} networks with MAC and broadcast components under the special traffic patterns presented in Sec.~\ref{sec:specialTraffic}.
\bthm For a directed Gaussian network composed of broadcast and multiple access channels, a simple separation strategy can achieve a rate,
\beqa \mc{R}_{\ach}(P) & \supseteq & {\mc{R}_{\cut} \left( \frac{P}{d} \right)} \quad \text{for BC Traffic}, \\
{R}^{\text{sum}}_{\ach}(P) & \supseteq & {R}^{\text{sum}}_{\cut} \left(\frac{P}{d} \right) \quad \text{for X Traffic}, \\
 {R}^{\text{sum}}_{\ach}(P)  & \supseteq & \frac{{R}^{\text{sum}}_{\cut} \left( \frac{P}{d} \right)}{2} \quad \text{for group-communication traffic}. \eeqa   \ethm
\bpf The proofs are based on the polymatroidal results Lemma~\ref{lem:directedBC}, Lemma~\ref{lem:directedX} and Lemma~\ref{lem:directedGroup}. The proof is very similar to the proof of Theorem~\ref{thm:bc_mac} and are therefore omitted. \epf

\subsection{Broadcast Erasure Networks with Commensurate Feedback \label{sec:Erasure}}
Broadcast erasure networks, in which there are broadcast but no superposition constraints, serve as high level models for communication in wireless networks. Unicast in broadcast erasure networks is well understood, for which it has been shown \cite{Erasure} that min-cut is achievable using a global linear network coding scheme (in  \cite{ErasureMédard}, it is shown that knowledge of erasure locations is not necessary at the destination) . It has also been shown that a separation scheme in which each broadcast erasure channel is coded for locally to create noiseless links does not perform very well. This is due to the fact that for each broadcast erasure channel the capacity region is far away from the  min-cut region. However, as shown in Sec.~\ref{sec:BEC}, by utilizing ACK feedback, the capacity region is enlarged to become closer to the min-cut region. Therefore we consider a network of broadcast erasure channel, where each channel has a feedback mechanism.

Consider a network which is composed of broadcast erasure channels, with an appropriate mechanism for feedback built into the network. In particular, we look for feedback that is commensurate; formally: a channel is said to have {\em commensurate feedback} if there are feedback links from the various receiving nodes to the transmitters with the same rate region as the cut-set bound for the forward channel.  The reciprocal nature of wireless channels from which the broadcast erasure channel is constructed naturally suggests a way of providing feedback links of commensurate strength. In Appendix~\ref{app:MAC_Erasure}, we look at one possible way of obtaining feedback links of commensurate strength as the forward links.

 From now on, we will assume that each erasure broadcast channel has commensurate feedback, without cognizance to how this particular rate region was obtained. For simplicity, we will assume that all broadcast erasure channels are symmetric and independent, i.e., each broadcast erasure channel has erasure independent probability $\epsilon$.

\bthm \label{thm:erasure_multiple_unicast} For the $k$ unicast problem in a network of erasure broadcast channels with commensurate feedback, a simple separation strategy can achieve a rate
\beqa \mc{R}^{\text{erasure}}_{\ach} \supseteq \frac{\mc{R}^{\text{erasure}}_{\cut}}{\bigO (\log k) \bigO (\log d_{\max})}, \eeqa
where $d_{\max}$ is the maximum number of users in any broadcast erasure channel.
\ethm

\bpf Consider the following separation strategy: even though the feedback links have a rate region given by $R^{\text{BC}}_{\text{cut}}$, we will restrict them to use the rate region $R^{\text{BC}}_{\text{ach,fb}}$ in order to preserve symmetry.  The feedback links are used for two distinct purposes:
\beit \item To provide Ack / Nack feedback. This feedback has an overhead of $1$-bit per packet which we treat as  negligible. This assumption makes sense especially when packet lengths are large.
\item To route flows on the reverse direction, since the Ack feedback overhead is assumed to be small, this will essentially occupy the whole capacity. We establish a bidirected network by using the feedback links for routing.
\eeit
Since we have the Ack / Nack feedback for each erasure broadcast channel, we can use the scheme of Lemma~\ref{lem:BEC_FB} to obtain the rate region $R^{\text{BEC}}_{\text{ach,fb}}$ with feedback.
This induces a bidirected polymatroidal network specified in the following manner, in which we can use flows to achieve a rate region $R^{\poly}_{\text{ach}}$. By Theorem~\ref{thm:bidirected}, we have that \beqa R^{\poly}_{\text{ach}} \supseteq \frac{\mc{R}^{\poly}_{\cut}}{\bigO (\log k)}. \label{eq:fb_sep_ach} \eeqa
By Lemma~\ref{lem:BEC_FB_Cut}, we have $\mc{R}^{\text{BEC}}_{\text{ach,fb}} \supseteq \frac{ \mc{R}^{\text{BEC}}_{\text{cut}}}{\bigO ( \log d_{\max})}$. Further, since cuts in the polymatroidal and the original network decompose into cuts for each channel, any cut-set in the polymatroidal network induced by the achievable scheme has a counter-part cut-set in the erasure network within a factor of $\bigO(\log d_{\max})$:
    \beqa \mc{R}^{\poly}_{\cut} = \frac{\mc{R}^{\text{erasure}}_{\cut}}{\bigO (\log d_{\max})}.  \label{eq:fb_sep_cut} \eeqa
 Now \eqref{eq:fb_sep_ach} and \eqref{eq:fb_sep_cut} together imply,
\beqa \mc{R}^{\poly}_{\ach} \supseteq \frac{\mc{R}^{\text{erasure}}_{\cut}}{\bigO (\log k) \bigO (\log d_{\max})}, \eeqa
which proves the desired result.
\epf

\subsubsection{Special Traffic Scenarios}

We now present results for {\em directed} networks with broadcast erasure channels under the special traffic patterns presented in Sec.~\ref{sec:specialTraffic}. Since the networks are directed, reciprocity is not needed; however we will continue to assume that the broadcast erasure channel has ACK feedback.

\bthm For a directed network composed of broadcast erasure channels with ACK feedback, a simple separation strategy can achieve a rate,
\beqa \mc{R}_{\ach} & \supseteq & \frac{\mc{R}_{\cut}}{\log (d_{\max}+1)} \quad \text{for BC Traffic}, \\
{R}^{\text{sum}}_{\ach} & \supseteq & \frac{{R}^{\text{sum}}_{\cut}}{\log (d_{\max}+1)} \quad \text{for X Traffic}, \\
 {R}^{\text{sum}}_{\ach}  & \supseteq & \frac{{R}^{\text{sum}}_{\cut}}{2 \log (d_{\max}+1)} \quad \text{for group-communication traffic}, \eeqa   where $d_{\max}$ is the maximum degree of the broadcast channel. \ethm

\bpf The proofs are similar to the proof of Theorem~\ref{thm:erasure_multiple_unicast} and are therefore omitted. \epf

\subsection{Gaussian Fast Fading Network}
We will now consider a general Gaussian network where broadcast and superposition can simultaneously occur, i.e., the network can contain interference components. Such a network is clearly more general than the $K$-user interference channel, and even this channel has not been well understood in the most general case ( the tutorial \cite{JafarTutorial} provides an excellent summary of the current understanding on this channel). However, in the presence of fast fading, the problem gets symmetrized considerably \cite{ErgodicIA}, and there is a reasonable understanding of this problem. Therefore we will resort to the fast fading model in this section. We will further assume that the fading distribution satisfies the assumptions in Sec.~\ref{sec:singleHop}.

While most of the existing literature is on single hop interference channels, multi-hop interference networks have been the focus of more recent work.  In particular, it has been shown in \cite{JafarChungDOFLayered} that the degrees of freedom of such fully-connected layered networks can be achieved using a non-separation scheme called opportunistic interference alignment. It has also been shown \cite{JafarChung2X2X2} that a separation architecture does not even achieve the DOF for simple networks, for example, the network with $2$ sources, $2$ relays and $2$ destinations. Our results offer a contrast: if we look to achieve within  an approximation factor of cut-set then a simple  separation strategy suffices, for all SNR.

\begin{figure}[htb]
\begin{center}
\vspace{0.1in}
\scalebox{0.5}{\includegraphics{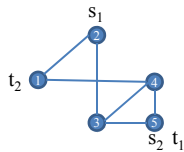}}
\end{center}
\caption{A Multiple-Unicast Wireless Network} \label{fig:nonLayered}
\end{figure}

\begin{figure}
  \centering
 \vspace{0.1in}
  \subfloat[Layered Network]{\label{fig:undirectedProblem}{{\scalebox{0.65}{\includegraphics{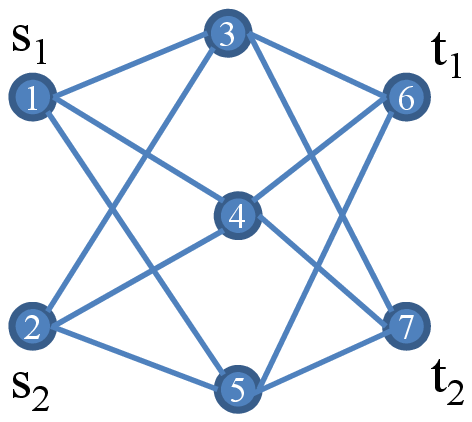}}}}}
  \hspace{0in} \subfloat[Cuts]{\label{fig:undirectedLayered}{ {\scalebox{0.65}{\includegraphics{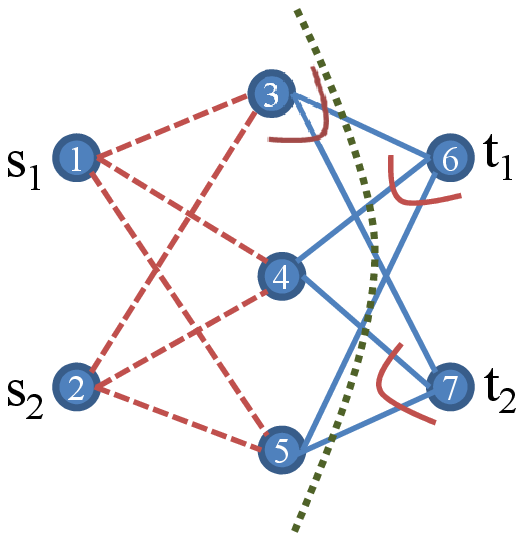}}}}}
  \caption{Layered Network and Cuts}
  \label{fig:Layered}
\end{figure}

For examples of networks considered here, we refer the reader to Fig.~\ref{fig:nonLayered} for a non-layered example or the one in Fig.~\ref{fig:Layered} for a layered example.

\bthm \label{thm:bidirected_fading} For a bidirected ergodic wireless network with $k$ source destination pairs, the rate region given by \beqa \mc{R}^g ( P)  \supseteq \frac{\mc{R}^g_{\cut} \left( \frac{P}{bd^3} \right)}{\bigO (\log k)}, \eeqa
is achievable using a separation strategy, where $d$ is the maximum degree of any node and $b:=\frac{e^{-\mb{E}( \log  |h|^2)}}{2}$ is a constant depending on the fading distribution. If the fading is complex Gaussian, $b \approx 0.86 $.
\ethm

\bpf
We will now show a scheme by which we can convert the bidirected ergodic wireless network into a bidirected polymatroidal network. We will view one snapshot of transmission in the network as a transmission in a bipartite graph, from the set of nodes $V^l = V$ on the left side to the set of nodes $V^r$ on the right side. Node $u$ on the left is connected to node $v$ on the right with channel coefficient $h_{uv}$ from the original network. The nodes have infinite memory, so each node is connected to itself with infinite capacity. We view the obtained network as a single hop network and use a scheme for the single hop (described in Sec.~\ref{sec:singleHop}). The achievable rates for this single hop network are given by polymatroidal constraints, and therefore we obtain a polymatroidal network represented by a bi-partite graph with $2V$ nodes. This process can now be {\em reverted}, i.e., we can go from a layered representation with $2V$ nodes to a non-layered representation with $V$ nodes. Thus we obtain a polymatroidal network
 with $V$ nodes. There are edges $E$ similar to the original graph, however the capacities are constrained according to the polymatroidal constraint:
\beqa \sum_{u \in \In{v}} R_{uv} \leq \frac{1}{2} C(2P) \quad \forall v \in V \\
\sum_{u \in \Out{v}} R_{vu} \leq \frac{1}{2} C(2P) \quad \forall u \in V . \eeqa
This is now a bidirected polymatroidal network since the polymatroidal constraint is symmetric for the incoming and the outgoing edges at any given node. Now we perform routing over this bidirected polymatroidal network. Now the flow and cut on this network are related by:
\beqa \mc{R}_{\ach}^{\g} = \mc{R}_{\ach}^{\poly} \supseteq  \frac{\mc{R}_{\cut}^{\poly}}{ \bigO (\log k)}. \label{eq:poly_flow_cut_int} \eeqa

If we can relate the cut-set bound on the polymatroidal network and the cut-set bound on the original Gaussian network,  we can get the desired result. This is done in the following lemma:
\blem \label{lem:poly_g_cut}
\beqa \mc{R}^{\poly}_{\cut}(P) \supseteq \frac{1}{2}  \mc{R}^{\org}_{\cut} \left( \frac{P}{b d^3} \right). \label{eq:blp3} \eeqa \elem

\bpf See Appendix \ref{app:poly_g_cut} \epf

Now, \eqref{eq:poly_flow_cut_int} and \eqref{eq:blp3} implies that
  \beqa \mc{R}_{\ach}^{\org} \supseteq \frac{\mc{R}^{\org}_{\cut} \left( \frac{P}{b d^3} \right)}{ \bigO (\log k)}, \eeqa which completes the proof of the theorem. \epf

\subsubsection{Multi-Antenna Nodes \label{sec:multiAntenna_Interference}}
We can prove a result similar to the one in Theorem~\ref{thm:bidirected_fading} even when there are multiple antennas at the nodes.

\bthm \label{thm:bidirected_multi} For a bidirected ergodic wireless network with multiple antenna nodes, the rate region for $k$-unicast is given by
\beqa \mc{R} ( P)  \supseteq \frac{\mc{R}_{\cut} \left( \frac{P}{bd^3} \right)}{\bigO (\log k)}, \eeqa
is achievable using a separation strategy, where $d$ is the maximum number of antennas that can communicate with any given antenna and $b:=\frac{e^{-\mb{E}( \log  |h|^2)}}{2}$ is a constant depending on the fading distribution.
\ethm

\bpf The proof of the theorem is very similar to that of Theorem~\ref{thm:bidirected_fading} except that when there are multiple antennas, each antenna is treated as a separate node with infinite capacity wireline links between antennas of the same node. This scheme can be shown to achieve the cut-set of this new network to within the approximation factor. We observe that any finite cut in this new network will partition the antennas in such a way that all antennas corresponding to the same original node will lie on the same side of the partition, since otherwise the value of the cut will become infinite. Therefore the cut-set in the new network and the original network have the same value and this completes the proof of the theorem.
\epf

\subsubsection{Special Traffic Scenarios}
We now present results for {\em directed} fast fading networks under the special traffic patterns presented in Sec.~\ref{sec:specialTraffic}. Since the network is directed, reciprocity will not be necessary to prove this result.
\bthm For a directed fast fading Gaussian network with multiple antenna nodes, a simple separation strategy can achieve a rate,
\beqa \mc{R}_{\ach}(P) & \supseteq & \frac{\mc{R}_{\cut} \lbr \frac{P}{bd^3} \rbr }{2} \quad \text{for BC Traffic}, \\
{R}^{\text{sum}}_{\ach}(P) & \supseteq & \frac{{R}^{\text{sum}}_{\cut} \lbr \frac{P}{bd^3} \rbr}{2} \quad \text{for X Traffic}, \\
 {R}^{\text{sum}}_{\ach}(P)  & \supseteq & \frac{{R}^{\text{sum}}_{\cut} \lbr \frac{P}{bd^3} \rbr }{4} \quad \text{for group-communication traffic}, \eeqa where $d$ is the maximum number of antennas that can communicate with any given antenna and $b:=\frac{e^{-\mb{E}( \log  |h|^2)}}{2}$ is a constant depending on the fading distribution.  \ethm
\bpf The proofs are similar to the proof of Theorem~\ref{thm:bidirected_fading} and are therefore omitted. Here, the factor loss of $2$ is present for the BC scenario due the factor $2$ loss in the local physical layer scheme (see Theorem~\ref{thm:single_layer_X}). \epf

\subsubsection{Special Channel Model: Directed Layered Network}
\begin{figure}
  \centering
  \subfloat[Wireless Network]{\label{fig:directedLayered}{\scalebox{0.65}{\includegraphics{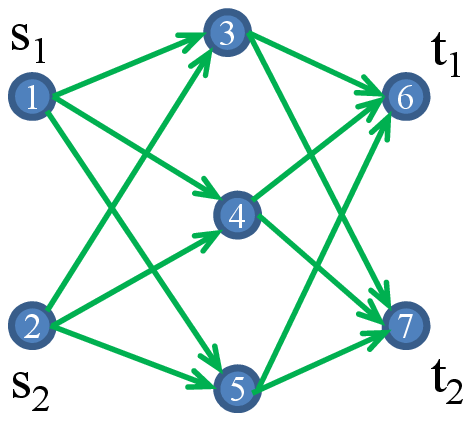}}}}
  \subfloat[Polymatroidal Network]{\label{fig:polymatroidDirectedLayered}{\scalebox{0.65}{\includegraphics{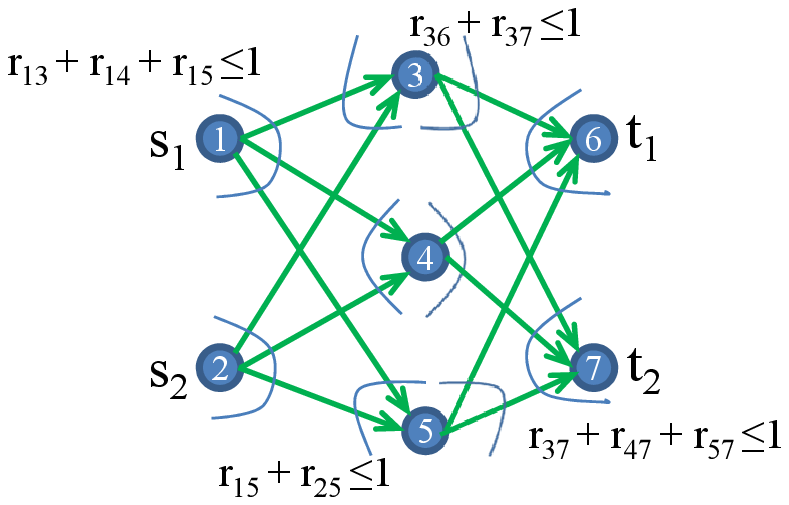}}}}
  \caption{Directed Layered F.C. Networks}
  \label{fig:directedLayeredTwo}
\end{figure}

In this section, we consider directed fully-connected (f.c.) layered networks. These are layered networks, which have connectivity between adjacent layers only in the forward direction, i.e., links are always between a node in $V_i$ to a node in $V_{i+1}$. Further for a fully-connected network, we assume that $(u,v) \in \mc{E} \quad \forall u \in V_{i}, v \in V_{i+1}$. Consider, for example, the network in Fig.~\ref{fig:directedLayered}.

%
%
%
%

\bthm \label{thm:directed_layered} For a directed fully-connected layered ergodic wireless network with $k$-distinct sources in the first layer having messages to $k$ distinct sinks in the last layer, the rate region given by
\beqa \mc{R} ( P)  \supseteq \frac{\mc{R}_{\cut}\lbr \frac{P}{bd^3} \rbr }{2}, \eeqa
is achievable using a separation strategy, where $d$ is the maximum degree of any node and  $b:=\frac{e^{-\mb{E}( \log  |h|^2)}}{2}$. \ethm

\bpf
The achievable scheme is again a local Phy scheme, i.e., for each hop in the layered network we use the strategy for X-channels described in Sec.~\ref{sec:singleHop}. Now we get a special type of directed polymatroidal network for which we can show that max-flow equals min-cut.

\bthm \label{thm:polymatroid_layered} For a $k$-unicast problem in a polymatroidal directed layered network in which all the node constraints are of the form
\beqa \sum_{u \in \In \{v \}} R_{uv} \leq 1, \quad \forall v, \\
\sum_{v \in \Out \{u \}} R_{uv} \leq 1, \quad \forall u, \eeqa
the rate region given by max-flow equals the min-cut for the $k$-unicast problem.
\ethm

\bpf See Appendix ~\ref{app:polymatroid_layered}. \epf

Now we can use Lemma~\ref{lem:poly_g_cut} to get:
\beqa \mc{R}^{\org}_{\ach}(P) = \mc{R}^{\poly}_{\ach}(P) = \mc{R}^{\poly}_{\cut}(P) \supseteq \frac{1}{2}  \mc{R}^{\org}_{\cut} \lbr \frac{P}{bd^3} \rbr , \eeqa
which implies the theorem.
\epf

\subsection{Fixed Gaussian Network}
Consider a general Gaussian network where broadcast and superposition can occur simultaneously, and where the channel coefficients are fixed but drawn from a continuous distribution. We studied a simple single-hop version of such a network in Sec.~\ref{sec:fixedSingleHop}. We will now show that the local scheme can be placed in a network context to get an almost-sure DOF characterization.

\bthm For a bidirected fixed multi-antenna Gaussian network with $k$ source destination pairs, the DOF given by \beqa \mc{D}_{\ach}   \supseteq \frac{\mc{D}_{\cut}}{\bigO (\log k)}, \eeqa
is achievable {\em almost surely} using a separation strategy. \ethm

\bpf The proof proceeds analogously to Theorem~\ref{thm:bidirected_multi}, but instead of using the local scheme in Theorem~\ref{thm:single_layer_X}, the scheme of Theorem~\ref{thm:single_layer_X_fixed} is used and therefore a DOF characterization is obtained. \epf

\subsubsection{Special Traffic Scenarios}
We now present results for {\em directed} Gaussian networks under the special traffic patterns presented in Sec.~\ref{sec:specialTraffic}. Since the network is directed, reciprocity will not be necessary to prove this result.
\bthm For a directed Gaussian network with each channel coefficient chosen from a continuous distribution, a simple separation strategy can achieve a DOF region,
\beqa \mc{D}_{\ach} & \supseteq & \frac{\mc{D}_{\cut}}{2} \quad \text{for BC Traffic}, \\
{D}^{\text{sum}}_{\ach} & \supseteq & \frac{{D}^{\text{sum}}_{\cut}}{2} \quad \text{for X Traffic}, \\
 {D}^{\text{sum}}_{\ach}  & \supseteq & \frac{{D}^{\text{sum}}_{\cut}}{4} \quad \text{for group-communication traffic}. \eeqa  \ethm

\subsection{Linear deterministic networks with MAC and Broadcast components}
A linear deterministic network composed of MAC and broadcast components is defined in the same way as the Gaussian network composed of MAC and broadcast components. The key difference is that the transmissions are over a finite field $\mb{F}_q$, there is no noise, and the channels between node $i$ and $j$ of color $c$ are matrices in general: $H_{ij}^c$.
The main result for linear deterministic networks with MAC and broadcast components is stated in the following theorem:

\bthm \label{thm:ld_bc_mac} For the $k$-unicast problem in linear deterministic network composed of broadcast and multiple access channels, the layered architecture can achieve a rate,
\beqa \mc{R}^{\text{ld}}_{\ach} \supseteq \frac{\mc{R}^{\text{ld}}_{\cut}}{\bigO (\log k)}.  \eeqa
\ethm

\bpf The proof of this theorem proceeds in a similar way as that of Theorem~\ref{thm:bc_mac}, the key difference being the fact that for linear deterministic networks, the cut-set bounds under product form and general distributions are the same, and therefore, there is no power scaling factor. \epf

\subsubsection{Special Traffic Scenarios}

We now present results for {\em directed} linear deterministic networks with MAC and broadcast components under the special traffic patterns presented in Sec.~\ref{sec:specialTraffic}.
\bthm For a directed linear deterministic network composed of broadcast and multiple access channels, a simple separation strategy can achieve a rate,
\beqa \mc{R}_{\ach} & \supseteq & {\mc{R}_{\cut}} \quad \text{for BC  Traffic}, \\
{R}^{\text{sum}}_{\ach} & \supseteq & {R}^{\text{sum}}_{\cut} \quad \text{for X Traffic}, \\
 {R}^{\text{sum}}_{\ach}  & \supseteq & \frac{{R}^{\text{sum}}_{\cut}}{2} \quad \text{for group-communication traffic}. \eeqa   \ethm

\bpf The proofs are similar to the proof of Theorem~\ref{thm:ld_bc_mac} and are therefore omitted. \epf

\subsection{Networks of Fast Fading MAC and Broadcast Channels with delayed CSIT \label{sec:FadingMacBc}}

We will now networks composed of fast fading MAC and Broadcast channels where the channel states are i.i.d. over antennas and time, and the CSI is available at the transmitting nodes after a delay. Schemes for each of these local channels was studied in Sec.~\ref{sec:Delayed_BC_MAC} and a degree of freedom characterization was obtained. Now, we will try to obtain the degree of freedom region for multiple unicast over a network of such channels.

Our main result for such networks is the following.
\bthm \label{thm:fading_bc_mac} For the $k$ unicast problem in Gaussian network composed of fading broadcast and multiple access channels with delayed feedback, a simple separation strategy can achieve a DOF,
\beqa \mc{D}_{\ach} \supseteq \frac{\mc{D}_{\cut}}{\bigO (\log k) \bigO (\log p_{\max})}, \eeqa
 where \beqa p_{\max} = \max_{\text{BC channels}} p(\text{BC channel}), \eeqa
and $p(\text{BC channel})$ is given by the minimum of number of transmit antennas and the total number of received antennas in the broadcast channel.
\ethm

\bpf The coding scheme is again a separation-based strategy: each component broadcast or multiple access channel with delayed feedback is coded for independently creating bit-pipes on which information is routed globally. The physical layer technique is the scheme for MIMO broadcast and MAC channels with delayed CSIT proposed in Sec.~\ref{sec:Delayed_BC_MAC}. The proof is very similar to that in  Sec.~\ref{sec:Erasure}.

 For the fading MIMO broadcast channel with delayed feedback, we can achieve the following DOF region $\mc{D}^{\text{BC}}_{\text{ach}}$ with feedback. For the fading MIMO multiple access channel, we can achieve the region given by $\mc{D}^{\text{MAC}}_{\text{ach}}$, however we will restrict ourselves to achieve the smaller rate region  $\mc{D}^{\text{BC}}_{\text{ach}}$ for the purpose of symmetry. This induces a bidirected polymatroidal network, in which we can use flows to achieve a rate region $R^{\poly}_{\text{ach}}$. By Theorem~\ref{thm:bidirected}, we have \beqa R^{\poly}_{\text{ach}} \supseteq \frac{\mc{R}^{\poly}_{\cut}}{\bigO (\log k)}. \eeqa
 By Lemma~\ref{lem:Fading_MIMO}, we have $\mc{D}^{\text{BC}}_{\text{ach}} \supseteq \frac{ \mc{D}^{\text{BC}}_{\text{cut}}}{\bigO ( \log p_{\max})}$ and by choice, $\mc{D}^{\text{MAC}}_{\text{ach}} = \frac{ \mc{D}^{\text{MAC}}_{\text{cut}}}{\bigO ( \log p_{\max})}$ . Further, since cuts in the polymatroidal and the original fading network decompose into cuts for each channel, any cut-set in the polymatroidal network induced by the achievable scheme has a counter-part cut-set in the erasure network within a factor of $\bigO(\log p_{\max})$:
 \beqa \mc{R}^{\poly}_{\cut} = \frac{\mc{D}^{\text{fading}}_{\cut}}{ \bigO (\log p_{\max}) } . \eeqa Also, we can achieve the same DOF in the original fading network as the rate in the polymatroidal network, i.e.,
\beqa \mc{R}^{\poly}_{\text{ach}} = \mc{D}^{\text{fading}}_{\text{ach}}. \eeqa
 This implies that
\beqa \mc{D}^{\text{fading}}_{\text{ach}} \supseteq \frac{\mc{D}^{\text{fading}}_{\cut}}{\bigO (\log^3 k) \bigO (\log p_{\max})}. \eeqa
This completes the proof of the theorem.
 \epf

\subsubsection{Special Traffic Scenarios}

We now present results for {\em directed} networks with MIMO broadcast and MAC channels with delayed feedback under the special traffic patterns presented in Sec.~\ref{sec:specialTraffic}. Since the networks are directed, reciprocity is not needed; however we will continue to assume that the broadcast channel gets delayed CSI feedback.

\bthm For a directed network composed of MIMO broadcast and MAC channels with delayed CSI feedback, a simple separation strategy can achieve a DOF region,
\beqa \mc{D}_{\ach} & \supseteq & \frac{\mc{D}_{\cut}}{\log (p_{\max}+1)} \quad \text{for BC Traffic}, \\
{D}^{\text{sum}}_{\ach} & \supseteq & \frac{{D}^{\text{sum}}_{\cut}}{\log (p_{\max}+1)} \quad \text{for X Traffic}, \\
 {D}^{\text{sum}}_{\ach}  & \supseteq & \frac{{D}^{\text{sum}}_{\cut}}{2 \log (p_{\max}+1)} \quad \text{for group-communication traffic}, \eeqa   where \beqa p_{\max} = \max_{\text{BC channels}} p(\text{BC channel}), \eeqa
and $p(\text{BC channel})$ is given by the minimum of number of transmit antennas and the total number of received antennas in the broadcast channel. \ethm

\bpf The proofs are similar to the proof of Theorem~\ref{thm:fading_bc_mac} and are therefore omitted. \epf

\subsection{Fading Linear Deterministic Network}
Consider a linear deterministic network of the form defined in Sec.~\ref{sec:fadingLD}, where each of the non-zero channel coefficients $H_{ij}(t)$ undergo i.i.d. fading with a uniform distribution on the non-zero elements. Then the local scheme of Sec.~\ref{sec:fadingLD} can be extended to a global network scheme.

\bthm For a bidirected linear deterministic network with $k$ source destination pairs, the rate region given by \beqa \mc{R}_{\ach}   \supseteq \frac{\mc{R}_{\cut}}{\bigO (\log k)}, \eeqa
is achievable using a separation strategy. \ethm

\bpf The proof proceeds analogously to Theorem~\ref{thm:bidirected_fading}, but instead of using the local scheme in Theorem~\ref{thm:single_layer_X}, the scheme of Theorem~\ref{thm:LD_X}. \epf

\subsubsection{Special Traffic Scenarios}
We now present results for {\em directed} fast fading linear deterministic networks under the special traffic patterns presented in Sec.~\ref{sec:specialTraffic}. Since the network is directed, reciprocity will not be necessary to prove this result.
\bthm For a directed fast fading linear deterministic network, a simple separation strategy can achieve a rate,
\beqa \mc{R}_{\ach} & \supseteq & \frac{\mc{R}_{\cut}}{2} \quad \text{for BC Traffic}, \\
{R}^{\text{sum}}_{\ach} & \supseteq & \frac{{R}^{\text{sum}}_{\cut}}{2} \quad \text{for X Traffic}, \\
 {R}^{\text{sum}}_{\ach}  & \supseteq & \frac{{R}^{\text{sum}}_{\cut}}{4} \quad \text{for group-communication traffic}. \eeqa \ethm

\subsection*{Acknowledgements} The authors would like to thank Chandra Chekuri and Adnan Raja for several useful discussions.

\pagebreak

\appendix

\section{Proof of Lemma~\ref{lem:cuts_poly_gauss} \label{app:cuts_poly_gauss}}

The rate region for cut-set under product distribution is given by:
\beqa \mc{R}^{\text{MAC}}_{\text{cut,product}}(P) =  \left \{R: \sum_{i \in S} R_i \leq \log(1 + \sum_{i \in S} |h_i|^2 P) \right \}. \eeqa

The rate region for cut-set under general distribution is given by:
\beqa \mc{R}^{\text{MAC}}_{\text{cut,general}}(P) =  \left \{R: \sum_{i \in S} R_i \leq \log(1 + (\sum_{i \in S} |h_i|)^2 P) \right \}. \eeqa

By the Cauchy-Schwarz inequality, we get
\beqa  \left (\sum_{i \in S} |h_i| \right )^2 P & \leq & \left ( \sum_{i \in S} |h_i|^2 \right ) dP, \eeqa
which in turn implies \beqa \mc{R}^{\text{MAC}}_{\text{cut,general}}(P) \subseteq \mc{R}^{\text{MAC}}_{\text{cut,product}}(d   P). \eeqa
We can similarly show that,
\beqa \mc{R}^{\text{BC}}_{\text{cut,general}}(P) = \mc{R}^{\text{MAC}}_{\text{cut,product}}(d   P) . \label{eq:bcmac_cut} \eeqa

Along with the equality in \eqref{eq:mac_rate_cut}, this implies that
\beqa \mc{R}^{\text{MAC}}_{\text{cut,general}}(P) \subseteq \mc{R}^{\text{MAC}}_{\text{ach}}(d   P). \eeqa
We similarly get, using \eqref{eq:bcmac_rate} and \eqref{eq:bcmac_cut},
\beqa \mc{R}^{\text{BC}}_{\text{cut,general}}(P) \subseteq \mc{R}^{\text{BC}}_{\text{ach}}(d   P). \eeqa

\section{Proof of Lemma~\ref{lem:BEC_No_FB} \label{app:BEC_No_FB}}

 Without feedback the capacity of this erasure broadcast channel can be easily found. This is because this erasure broadcast channel is stochastically degraded  \cite{CovTho}, and the capacity is given by
\beqa \lbr (R_1,...,R_d)|  \sum_{i=1,2,..,d} R_i \leq 1 - \epsilon \rbr. \eeqa
The rate region can be achieved by time sharing between the individual links.
We can compare this rate to the cut-set bound which is given by
\beqa R_{\text{cut}} = \lbr (R_1,...,R_D) | \sum_{i \in J} R_i \leq 1- \epsilon^{|J|} \quad \forall J \subseteq \{1,2,..,d\} \rbr. \eeqa
The ratio between the sum rate of the scheme and the cut-set bound is the factor
 \beqa \frac{1-\epsilon}{1-\epsilon^d}, \text{ which } & \rightarrow & \frac{1}{d}, \text{ as } \epsilon \rightarrow 1. \eeqa
 As expected, the time sharing region does not compare very favorably to the cut-set bound.

\section{Proof of Lemma~\ref{lem:BEC_FB_Cut} \label{app:BEC_FB_Cut}}
 Consider the rate region with feedback $\mc{R}_{\text{ach,fb}}$. We would like to know for what value of $A$ does $A \mc{R}_{\text{ach,fb}} \supseteq {\mc{R}_{\text{cut}}}$. Let us take a point in $\mc{R}_{\text{cut}}$, we would like to know, for what value of $A$ does this imply  $ \sum_{i =1,2,..,d} \frac{R_{i}}{1- \epsilon^i} \leq A$. This is equivalent to
\beqa A = \max \sum_{i=1}^n \frac{R_i}{1-\epsilon^i}, \eeqa
such that,
\beqa \sum_{i \in J} R_i \leq 1- \epsilon^{|J|} \quad \forall J \subseteq \{1,2,..,d\}. \eeqa
This is a linear optimization over a polymatroid and the optimal solution is given by the greedy algorithm \cite{Edmonds1},
\beqa (R_1,...,R_d) = (1-\epsilon, \epsilon-\epsilon^2,...,\epsilon^{n-1} - \epsilon^n), \eeqa
and the optimal value of the objective function is
\beqa \sum_{i=1}^n \frac{\epsilon^{i-1} - \epsilon^i}{1-\epsilon^i}. \eeqa

Lets examine the $i$-th term in this sum, substituting $x = \epsilon^{-1}$,
\beqa \frac{x-1}{x^i-1} = \frac{1}{1+x^1+..+x^{i-1}} \leq \frac{1}{i}. \eeqa
Therefore the sum is upper bounded by \beqa A \leq \sum_i \frac{1}{i} \leq \log d. \eeqa

\section{Feedback: Multiple Access Erasure Channel \label{app:MAC_Erasure}}
Consider a finite field multiple access erasure channel, where
\beqa y = \sum_{i=1}^d e_i x_i, \label{eq:FF_Mac} \eeqa
where $e_i$ are i.i.d. Bernouli with probability $1-\epsilon$.
In this channel, some of the transmitters' packets can get erased, and the received vector is the sum of those packets that did not get erased.

This multiple access channel is the dual of the broadcast erasure channel in the sense that the cut-set bound of the two channels are identical. This channel can be realized physically by having using computation code on the wireless channel, which computes the required linear combination. If all the channel coefficients in the wireless channel are good, then the combination $\sum_{i=1}^d X_i$ can be computed easily. However if one of the channel coefficient, say $h_j$, is in deep fade, then it may be algorithmically hard to compute this linear combination. Therefore, one way around this problem is to avoid having $X_j$ in the linear combination and instead compute $\sum_{i=1,2,...,d \ i \neq j} X_i$. This gives rise to the channel model in \eqref{eq:FF_Mac}.

The capacity of this multiple access channel is given by the cut-set bound (which is the same as the broadcast channel cut-set bound),
\beqa \mc{R}^{\text{MAC}} = \mc{R}^{\text{MAC}}_{\text{cut}} = \{ (R_1,...,R_D) | \sum_{i \in J} R_i \leq 1- \epsilon^{|J|} \quad \forall J  \}. \eeqa
Here $J \subseteq \{1,2,..,d\}$. We emphasize
 that the capacity region of this multiple access channel is equal to the cut-set bound of the erasure broadcast channel.

\section{Proof of Lemma~\ref{lem:Fading_MIMO} \label{app:fading_mimo}}

We will first consider the case of a network with $l$ transmit antennas and $m$ single-antenna receivers.
We can assume $l \leq m$, since if $l > m$, we can restrict ourselves to using $m$ transmit antennas, which leaves the cut unaltered. Therefore $p = l$.

We can choose any particular subset of $l$ receivers and use the strategy in Lemma~\ref{lem:AliTse} to achieve a DOF of $\frac{1}{\bigO( \log l)}$ for each receiver. We can time share between all possible subsets of size $l$ to achieve a certain DOF region. To compute the rate region achievable by this method, we use the following trick: Let the DOF tuple achieved be $\frac{1}{\bigO( \log l)} (r_1,...,r_m) $.  Let us construct a bi-partite graph with $l$ nodes on the left partition and $m$ nodes on the right partition and a complete graph connecting them. Each matching is equivalent to choosing a certain subset of the receivers (given by the set of right-partition nodes covered by the matching) and achieving DOF $1$ for the each of the receivers. The characteristic vector of a bipartite matching is given by $(x_{ij})_M$ such that $x_{ij}=1$ for edge $(i,j)$ in the matching $M$ and $x_{ij}=0$ otherwise. The convex hull of these characteristic vectors is given by
\beqa \mc{M} = \text{conv} \lbr (x_{ij})_M | M \text{ a matching } \rbr. \eeqa
For a bipartite graph, this is equivalent to the following polytope \cite{Schrijver}
\beqa \mc{P} = \lbr (x_{ij}) | x_{ij} \geq 0 \quad \forall i,j,  \ \sum_{j} x_{ij} \leq 1 \quad \forall i \ \sum_i x_{ij} \leq 1 \quad \forall j \rbr. \eeqa

The DOF $d_j$ is given by \beqa d_j = \frac{1}{\bigO( \log l)} \sum_i x_{ij}. \eeqa

Now consider the following polytope,
\beqa \mc{D}_{\text{ach}} = \left \{(d_j) | d_j \geq 0 \quad \forall j, \ d_j \leq \frac{1}{\bigO( \log l)}, \  \sum_{j} d_j \leq \frac{l}{\bigO (\log l)} \right \}.  \eeqa

We can show that $\mc{D}_{\text{ach}}$ is equivalent to $\mc{P}$ by using the mapping
\beqa \psi: \mc{D}_{\text{ach}} & \rightarrow & \mc{P} \\
\psi \{ (d_j)\} & = & (x_{ij}) : x_{ij} = \bigO( \log l)  \frac{d_j} {l}, \eeqa
and the mapping
\beqa \zeta: \mc{P} & \rightarrow & \mc{D}_{\text{ach}} \\
\zeta \{ (x_{ij})\} & = & (d_j): d_j = \frac{1}{\bigO( \log l)} \sum_i x_{ij}. \eeqa

Thus the region $\mc{D}_{\text{ach}}$ is achievable. Also the cut-set bound is given by
\beqa \mc{D}_{\text{cut}} = \lbr (d_j) | d_j \geq 0 \quad \forall j, \ d_j \leq {1}, \  \sum_{j} d_j \leq {l} \rbr. \eeqa

This implies that
\beqa \mc{D}_{\text{ach}} = \frac{\mc{D}_{\text{cut}}}{\bigO( \log l)}, \eeqa
which completes the proof of the single antenna receiver case.

For the multi-antenna receiver case, we will treat it as being composed of many single antenna receivers each of which are receiving independent information and then sum up the rates. This proof will extend to this case to get the desired result.


\section{Proof of Lemma~\ref{lem:poly_g_cut} \label{app:poly_g_cut}}

 Let us start with a cut on the polymatroidal network. A cut on the polymatroidal network specifies a vertex partition $\Omega$ and along with it, specifies how to compute the value of the cut, by specifying which edges to group together for the sub-modular constraints. For example, consider the cut in Fig.~\ref{fig:undirectedLayered}, it features the vertex partition and also specifies how to group edges to get an upper bound.

From the given partition we need to construct a cut on the Gaussian network. The polymatroidal cuts specify which of the edges need to be grouped together. By bounding the polymatroidal network in this manner, each edge is involved in {\em either} a broadcast constraint or a superposition constraint. This is in contrast to the Gaussian case, where each cut has a certain value, and there is no sense in which edges are assigned to broadcast or superposition constraint.

The key idea to connect these two cuts is the idea of {\em decoupling} the constraints in the Gaussian channel:
 \beit \item In the Gaussian channel, if broadcast constraint is not active for a given edge, then the edge only participates in the superposition constraint and vice versa.
 \item While this is not true as it is in the Gaussian network, we can obtain an upper bound network where certain broadcast or superposition constraints are no longer active.  This process of obtaining upper bound network where a ceratin constraint is not active is called decoupling.
 \item Decoupling a {\em broadcast constraint} is easy, because a network in which edges are not involved in a broadcast constraint can only, in general, do better than a network where there is a broadcast constraint on the edges.
 \item  Decoupling the {\em superposition constraint} requires a bit more work, this can be illustrated using the following example. Suppose two edges $e_1$ and $e_2$ are involved in the superposition constraint in the following manner:
\beqa y = x_1 + x_2 + z. \eeqa
Then if we have construct another channel in which the edges $e_1$ and $e_2$ are not involved in a broadcast constraint, then we have,
\beqa y_i = x_i + z_i, i = 1,2, \eeqa
this channel can emulate any scheme in the original channel if $\text{Var}(z_i) = \frac{\text{Var}(z)}{2}$. If this condition is satisfied, then we can add up $y_1 + y_2$ to get $x_1 + x_2 + (z_1 + z_2)$, which is statistically equivalent to the original channel. Therefore to decouple the superposition constraint involving $d$ variables, we need to reduce the variance of the noise by $d$, the degree of the superposition constraint, or equivalently, increase the signal power by a factor $d$.

\item Therefore we can decouple all required broadcast and superposition constraints, if the power is increased by a factor of $d$, which is the maximum degree of any node.

     \eeit

Thus given any cut in the polymatroidal network along with the assignments of the edges to broadcast or superposition constraints, we can obtain a similar cut in the Gaussian network by decoupling the constraints which are not active in the polymatroidal cut. This incurs a power penalty factor of $d$, thus as far as the outer bound is concerned, we can assume that each node has power $dP$ instead of $P$. The network thus obtained is made of MAC and broadcast channels. In this network, every cut decomposes into the sum of MAC and BC cuts. A MAC cut with $d$ nodes when evaluated under general distribution on the input, is of the form,
\beqa \sum_i {R}_{ij} & \leq & \mb{E} \log \lpr 1+ (\sum_{i} |h_{ij}|)^2 d P \rpr \\
& \leq & \mb{E} \log \lpr 1+ \sum_{i} |h_{ij}|^2 d^2 P \rpr \quad \text{by Cauchy-Schwartz inequality} \\
& \leq &  \log \lpr 1+ \mb{E} (\sum_{i} |h_{ij}|^2) d^2  P \rpr \quad \text{by Jensen}\\
& = & \log \lpr 1+ d^3 P \rpr \\
& \leq  & \mb{E} \lbr \log \lpr 1 + a d^3 P |h|^2 \rpr \rbr = C(a d^3 P)  \eeqa
The last step follows because,
\beqa \mb{E} \lbr \log \lpr 1 + c |h|^2 \rpr \rbr & = &  \mb{E} \lbr \log \lpr 1 + c e^{\log  |h|^2} \rpr \rbr \\
& \geq & \log \lpr 1 + c e^{\mb{E}( \log  |h|^2)} \rpr.  \eeqa

Here $a := e^{-\mb{E}( \log  |h|^2)}$ is finite for the fading distribution, by assumption (it is equal to . Thus, the cutset bound for the original network implies that \beqa \sum_i {R}_{ij} & \leq & C \lpr a d^3 P \rpr, \eeqa
whereas the corresponding cutset bound for the polymatroidal network is of the form,
\beqa \sum_i {R}_{ij} & \leq & \frac{1}{2} C(2 P). \eeqa

We get,
\beqa \mc{R}^{\poly}_{\cut} \supseteq \frac{1}{2} \mc{R}^{\org}_{\cut} \lpr \frac{P}{b d^3} \rpr, \eeqa
where $b:=\frac{a}{2} = \frac{e^{\mb{E}( \log  |h|^2)}}{2}$ is a constant depending on the fading distribution. For $h$ distributed as complex gaussian, $b \approx 0.86$.

\section{Proof of Theorem~\ref{thm:polymatroid_layered} \label{app:polymatroid_layered}}

The $k$ sources are in the layer $V_0$, and the $k$ destinations are in the layer $V_{L+1}$. Let the number of nodes in the $i$-th layer be $n_i$.

\subsubsection{Max-Flow Rate}

Let $R_i$ be the rate between the $i$-th source destination pair. We will route the flow in a symmetric manner, where the incoming flow is divided equally among all the edges going out of a node. We will compute constraints on the rate region achievable by this strategy.

In the first hop, all edges going out of the $i$-th source will carry a flow of value $\frac{R_i}{n_1}$.
The constraint imposed by the edges going out of the source is given by
\beqa \frac{R_i}{n_1} n_1 \leq 1 \ \iff R_i  \leq 1, \quad \forall i = 1,2,..,k. \eeqa
The constraint imposed by the edges coming into the nodes of the first hop are given by
\beqa \frac{\sum_i R_i}{k n_1} k  \leq 1 \iff  \sum_i R_i \leq n_1.  \eeqa

The total flow carried by all the nodes in any given layer equals $\sum_i R_i$, and each node in layer $l$ carries a flow of $\frac{R_i}{n_l}$ corresponding to flow $i$.

In the $l-th$ hop connecting layers $V_{l-1}$ and $V_l$, each edge carries a flow of value $\frac{R_i}{n_l n_{l+1}}$ corresponding to source $i$, which yields a total flow of value $\frac{\sum_i R_i}{n_l n_{l+1}}$ for each edge.
The outgoing constraints on layer $l-1$ yield,
\beqa \frac{\sum_i R_i}{n_l n_{l+1}} n_{l+1} \leq 1 \ \iff \sum_i R_i  \leq n_l, \quad \forall i = 1,2,..,k. \eeqa
The incoming constraints on layer $l$ yield,
\beqa \frac{\sum_i R_i}{n_l n_{l+1}} n_{l} \leq 1 \ \iff \sum_i R_i  \leq n_{l+1}, \quad \forall i = 1,2,..,k. \eeqa

In the final ($L+1$-th) hop also, there are constraints similar to layer $1$. In particular, the outgoing constraints on layer $l$ yield,
\beqa \frac{\sum_i R_i}{n_L k} k \leq 1 \iff  \sum_i R_L  \leq n_L.  \eeqa
and the incoming constraints on the destination layer yield,
\beqa \frac{R_i}{n_L} n_L \leq 1 \ \iff R_i  \leq 1, \quad \forall i = 1,2,..,k. \eeqa

Thus a rate pair $(R_1,...,R_k)$ is achievable by routing iff
\beqa \sum_i R_i \leq \min (n_l, n_{l+1}) \quad \forall l = 0,1,...,L. \label{eq:flow1} \\
R_i \leq 1 \ \  \quad \forall i = 1,2,...,k. \label{eq:flow2} \eeqa

\subsubsection{Cut-set Region}
We can easily write the following constraints, which are a subset of the cut-set bounds.

Corresponding to the cut separating $\Omega = \cup_{i=0}^l V_i$ and $\Omega^c$, the following constraint can be written,
\beqa \sum_i R_i \leq \min (n_l, n_{l+1}) \quad \forall l = 0,1,...,L. \label{eq:cut1} \eeqa
Corresponding to the cut given by $\Omega = {s_i}$, we can get the following constraint
\beqa  R_i \leq 1 \quad \forall i = 1,...,k. \ \label{eq:cut2} \eeqa

Comparing \eqref{eq:flow1}, \eqref{eq:flow2} and \eqref{eq:cut1}, \eqref{eq:cut2}, we can deduce that any rate tuple that satisfies the cut-set region will lie in the rate region achieved by the flow. Thus the rate region corresponding to max-flow equals the rate region corresponding to cut-set region.

For example, Fig.~\ref{fig:polymatroidDirectedLayered} denotes the directed layered polymatroidal network obtained from the network in Fig.~\ref{fig:directedLayered}. Every node basically constrains the total inflow and outflow to be lesser than $1$.

\end{document}